\documentclass[prd,12pt]{article}
 \pdfoutput=1

\usepackage{amsmath,amssymb,graphicx,multirow,xspace,slashed}
\usepackage[colorlinks=true,urlcolor=blue,anchorcolor=blue,citecolor=blue,filecolor=blue,linkcolor=blue,menucolor=blue,pagecolor=blue]{hyperref}
\usepackage[compress,numbers]{natbib}
\usepackage{subfigure, placeins}
\usepackage{booktabs}
\usepackage{blindtext, rotating}
\usepackage{afterpage}
\usepackage{enumitem}
\usepackage{ marvosym }
\usepackage{authblk}
\usepackage{verbatim} 
\usepackage{soul} 
\usepackage[normalem]{ulem}

\usepackage{floatrow}
\newfloatcommand{capbtabbox}{table}[][\FBwidth]

\usepackage[font=footnotesize,labelfont=bf]{caption}

\usepackage{lineno}

\allowdisplaybreaks

\long\def\symbolfootnote[#1]#2{\begingroup%
\def\thefootnote{\fnsymbol{footnote}}\footnote[#1]{#2}\endgroup}

\newcommand{\newc}{\newcommand}
\newc{\gsim}{\lower.7ex\hbox{$\;\stackrel{\textstyle>}{\sim}\;$}}
\newc{\lsim}{\lower.7ex\hbox{$\;\stackrel{\textstyle<}{\sim}\;$}}
\newc{\gev}{\,{\rm GeV}}
\newc{\mev}{\,{\rm MeV}}
\newc{\ev}{\,{\rm eV}}
\newc{\kev}{\,{\rm keV}}
\newc{\tev}{\,{\rm TeV}}

\newc{\mz}{M_Z}
\newc{\mpl}{M_*}
\newc{\mw}{m_{\rm weak}}
\newc{\nr}[1]{N^c_R{}_{#1}}

\def\beq{\begin{equation}}
\def\eeq{\end{equation}}
\newcommand{\bea}{\begin{eqnarray}\begin{aligned}}
\newcommand{\eea}{\end{aligned}\end{eqnarray}}
\def\bitem{\begin{itemize}}
\def\eitem{\end{itemize}}
 \numberwithin{equation}{section}

\newcommand\fverb{\setbox\fverbbox=\hbox\bgroup\verb}

\newbox\fverbbox

\usepackage[usenames,dvipsnames]{xcolor}

\begin{document}

\baselineskip 0.6cm

\begin{titlepage}

\thispagestyle{empty}

\begin{center}

\vskip 1cm

{\Large\bf Cornering Natural SUSY at LHC Run II }\vskip0.2cm{\Large\bf and Beyond}

\vskip 1.0cm
{\large Matthew~R.~Buckley, David Feld, Sebastian Macaluso,\\\vskip0.2cm Angelo Monteux and David Shih }
\vskip 1.0cm
{\it NHETC, Dept.~of Physics and Astronomy\\ Rutgers, The State University of NJ \\ Piscataway, NJ 08854 USA} \\
\vskip 2.0cm

\end{center}

\begin{abstract}

We derive the latest constraints on various simplified models of natural SUSY with light higgsinos, stops and gluinos, using a detailed and comprehensive reinterpretation of the most recent 13 TeV ATLAS and CMS searches with $\sim 15$~fb$^{-1}$ of data. We discuss the implications of these constraints for fine-tuning of the electroweak scale. While the most ``vanilla'' version of SUSY (the MSSM with $R$-parity and flavor-degenerate sfermions) with 10\% fine-tuning is ruled out by the current constraints, models with decoupled valence squarks or reduced missing energy can still be fully natural. However, in all of these models, the mediation scale must be extremely low ($<100$~TeV). We conclude by considering the prospects for the high-luminosity LHC era, where we expect the current limits on particle masses to improve by up to $\sim 1$~TeV, and discuss further model-building directions for natural SUSY that are motivated by this work.

\end{abstract}

\end{titlepage}

\setcounter{page}{1}

\section{Introduction and Summary}

Recently,  at the ICHEP 2016 Conference \cite{ICHEP16}, the ATLAS and CMS collaborations presented the results of the search for new physics using the first $\sim 15$~fb$^{-1}$ at 13 TeV \cite{ATLAStwiki,CMStwiki}. This represents a significant milestone for the LHC: with this dataset the sensitivity to new physics at the energy frontier begins to truly surpass that achieved at Run I.  Now is therefore the perfect time to assess the implications of these results for well-motivated models of new physics such as supersymmetry (SUSY).

Weak-scale SUSY has long occupied a central place in the theoretical expectations for the LHC, as the addition of superpartners to the Standard Model (SM) particles at or near the scale of electroweak symmetry breaking stabilizes the Higgs mass and solves the hierarchy problem (for a review and original references, see e.g.~\cite{Martin:1997ns}).  Given that the superpartners must be heavier than their Standard Model counterparts, the supersymmetric cancellation of loops protecting the Higgs mass cannot be perfect. The heavier the superpartners, the more finely tuned the original bare mass must be against the loop contributions. If we require the theory be fully ``natural'' -- that the level of fine-tuning be less than some fixed amount, taken in this work to be the arbitrary threshold of 10\% -- then an upper limit can be derived on the mass of the Standard Model SUSY partners \cite{Barbieri:1987fn}. 
For a recent review on naturalness in SUSY with many original references, see e.g.\ \cite{Feng:2013pwa}.

While specific realizations of SUSY can have a wide variety of predictions for the spectrum of the superpartner masses, basic requirements of naturalness which hold in a wide class of models include a light higgsino (which directly sets the Higgs mass squared parameter at tree-level), relatively light stop squarks (as the top has an ${\cal O}(1)$ Yukawa, leading to large one-loop-corrections to the Higgs), and a relatively light gluino (which corrects the mass of the stop itself, yielding a two-loop correction to the Higgs)  
\cite{Dimopoulos:1995mi,Cohen:1996vb}. 

As the higgsino is color-neutral, it is difficult to produce directly and detect at the LHC. However, if the gluino and stop are relatively light as required by naturalness, they will be copiously produced at the LHC. As they cascade decay down to the light higgsinos, they will typically yield at least one or more of the following signatures   \cite{Evans:2013jna}: 
\begin{itemize}
\item Significant missing transverse momentum (MET)
\item Top quarks
\item High object multiplicity
\end{itemize}

The purpose of this paper is to investigate the parameter space of natural SUSY that is still allowed by the latest LHC searches. So far, no search has turned up definitive evidence for new physics, and the null results are phrased in terms of limits on a small set of ``simplified models''. 
In order to carry these limits over to more general scenarios, e.g.\ those motivated by natural SUSY, a detailed reinterpretation (``recasting'') of the LHC searches is required. In this work, we have performed a  comprehensive recasting of the 13 TeV post-ICHEP analyses which are most relevant for natural SUSY (i.e.\ target the signatures listed above), see Table~\ref{tab:searches} for a complete list. (Previous work that reinterpreted recent 13 TeV LHC results includes \cite{Kowalska:2016ent,Han:2016xet}.)
For more details and the validation of our simulations by comparison with the official ATLAS and CMS limits, see Section~\ref{sec:recast} and Appendix~\ref{app:recast}.

Using the recasted searches, we will explore the parameter space of natural SUSY, using a carefully chosen set of representative simplified models. Our philosophy here will be similar to that of \cite{Evans:2013jna}: we work purely bottom up, motivated to find the most conservative limits on gluinos, stops and higgsinos. In all the models we consider, the lightest MSSM sparticle is the higgsino, with $\mu\le 300$~GeV as suggested by naturalness. 
 However, apart from this assumption, we allow ourselves a great deal freedom in the simplified models.\footnote{Since this paper is focused on the implications of the latest LHC direct searches for natural SUSY, we will not require our simplified models to  raise the  Higgs mass to 125~GeV, instead remaining agnostic as to the source of this mass. As is well understood, if the SUSY Higgs mass corrections arise only from the MSSM stop squark loop, the level of tuning is at the few percent level or worse (see e.g.~\cite{Casas:2014eca} for a recent detailed discussion and references).}

First, as an essential baseline model, we will consider ``vanilla SUSY'' 
-- the 
 minimal supersymmetric Standard Model (MSSM) with $R$-parity conservation and flavor-degenerate sfermions. Next we will examine simplified models of natural SUSY that alleviate the latest LHC constraints by either reducing the signal cross section or by reducing the signal acceptance.
We will consider the ``effective SUSY'' scenario \cite{Dimopoulos:1995mi,Cohen:1996vb} where the 1st/2nd generation squarks are decoupled. Decoupling the valence squarks in particular reduces the total SUSY cross section by factors of ${\cal O}(10)$ or more in the region of the mass plane near the current LHC limits.  We also consider two scenarios that trade MET for jets: baryonic $R$-parity violating (RPV) decays of the higgsino (see e.g.~\cite{Barbier:2004ez} for a review and original references); and a hidden-valley (HV) \cite{Strassler:2006qa,Strassler:2006im} scenario inspired by Stealth SUSY  \cite{Fan:2011yu,Fan:2012jf}. By trading MET for jets, the signal becomes more difficult to distinguish from QCD, thereby significantly degrading the acceptance.

By examining the LHC limits on these simplified models, we will attempt to draw more general conclusions on viable directions for SUSY models post-ICHEP.  For each simplified model, we will overlay the current LHC limits in the gluino/stop mass plane together with the $\Delta\leq10$ natural regions, for different choices of the higgsino mass $\mu$ 
and the messenger scale $\Lambda$. 
Here $\Delta$ is derived from the Barbieri-Giudice fine-tuning measure \cite{Barbieri:1987fn} with respect to a soft SUSY-breaking parameter $M$, 
reformulated in terms of $m_h$ instead of $m_Z$ \cite{Kitano:2006gv} in order to better take into account the effect of radiative corrections to the Higgs quartic:
\beq
\Delta_{M^2} = {\partial \log m_h^2\over \partial \log M^2}
\eeq
When multiple sources of tuning are present, we take the maximum tuning as our  measure, $\Delta=\max_{\{M_i\}}\Delta_{M_i^2}$.

For the calculation of $\Delta_{M^2}$,  it has been conventional in much of the literature to work in the leading-log (LL) approximation (see however \cite{Essig:2007kh,Arvanitaki:2013yja,Casas:2014eca}  for notable exceptions). There the  quadratic sensitivity of the Higgs mass-squared parameter to the higgsino, stop and gluino soft masses arises at tree level, one-loop and two-loops respectively: 
\begin{itemize}

\item{} Higgsinos:
\beq\label{deltamhusqhiggsino}
\delta m_{H}^2 = \mu^2
\eeq
\item{} Stops:
\beq\label{deltamhusqstop}
\delta m_{H}^2\sim -{3\over8\pi^2}y_t^2 m_{\rm stop}^2 \log {\Lambda\over Q}
\eeq
\item{ } Gluinos:
\beq\label{deltamhusqgluino}
\delta m_{H}^2\sim -{g_3^2y_t^2\over 4\pi^4}|M_3|^2\left(\log{\Lambda\over Q}\right)^2 
\eeq
\end{itemize}
Here $\Lambda$ is the messenger scale of SUSY breaking, and $Q$ is the IR scale, conventionally taken to be 1~TeV in many works (see e.g.~\cite{Papucci:2011wy,Casas:2014eca}). Aside from the naturalness bounds this yields on the higgsino, stop and gluino masses, these LL formulas also demonstrate that the tuning is worsened as the messenger scale is raised. Natural SUSY theories greatly prefer lower values of $\Lambda$.

In this work, we go beyond the leading-log approximation and include a number of important higher-order effects,  including the full two-loop RGEs, one and two-loop threshold corrections to stop and gluino masses and threshold corrections to the Higgs potential. 
A detailed description of these effects  (and original references) will appear in a companion paper \cite{Buckley:2016tbs}. 
Here we will summarize the main idea: we translate the tuning bounds on the UV mass parameters $M^2$ which enter into the Barbieri-Giudice measure into upper limits on pole masses at the IR scale. The physical pole masses are what the LHC sets limits on. Surprisingly, the full set of differences between UV and IR parameters in tuning calculations has largely been neglected in the literature so far, but we find that they have several crucial consequences.

First, they are numerically important and they can raise the tuning bounds on sparticles by  $\mathcal{O}(1)$ factors. Second, including these higher-order corrections makes $\Delta_{M_3^2}$ ($\Delta_{m_{Q_3}^2}$) dependent on the stop (gluino) mass. Large gluino masses significantly raise the stop IR mass through the RG, while large stop masses contribute non-negligible threshold corrections to the gluino pole mass. Therefore in a natural spectrum, the stop and gluino mass are actually correlated: a heavy stop implies a heavy gluino, and vice versa. Perhaps it is counterintuitive, but the fact that we have seen neither the stop nor the gluino may be {\em more} consistent with natural SUSY than the discovery of one and not the other. In any event, this means that the $\Delta\leq10$ natural region is not simply a rectangle in the gluino/stop mass plane, but instead turns out to be wedge-shaped. 
Finally, the higher-order corrections include effects from the 1st/2nd generations that become very important in effective SUSY scenarios. At one-loop, heavy 1st/2nd generation squarks appreciably lift the gluino pole mass, which helps to relax the tuning bound on the gluino. At two-loops, the RGEs drive the stop mass lower in the IR, which can strengthen the tuning bound on the stops. See \cite{Buckley:2016tbs} for more details.

We will see that vanilla SUSY is  strongly constrained by the current searches and cannot be natural at the 10\% level for any choice of the messenger scale -- this was true already after Run I.
With either Effective SUSY or models that trade MET for jets, the LHC limits are greatly reduced, but still eliminate most of the parameter space with $\Delta\leq10$, except at the very lowest messenger scales $\Lambda\lesssim 20$~TeV. Finally, we consider the combination of Effective SUSY with RPV, and show that significant natural parameter space still remains at $\Lambda\lesssim$~100 TeV.  As many of models put forward as alternative solutions to the Hierarchy Problem {\em start} with at least 10\% tuning \cite{Chacko:2005pe,Burdman:2014zta,Bellazzini:2014yua}, the continued survival of natural SUSY serves as a reminder that -- despite the lack of discovery -- supersymmetry remains one of the least tuned solutions for physics beyond the Standard Model.\footnote{We note that there have been papers in the literature, even during Run I, that have claimed SUSY is at least percent-level tuned in all circumstances. Obviously, given the content of this paper, we believe these claims were  premature. A clear point of reference is with the work of  \cite{Arvanitaki:2013yja}: although we consider a rather similar set of simplified models,  we come to completely different conclusions about the tuning. A more in-depth comparison reveals the sources of the discrepancy. Aside from using a different measure of fine-tuning (summing in quadrature and multiplying tunings, vs.\ taking the max of the EW tuning), the main difference is that in \cite{Arvanitaki:2013yja}, broad conclusions about fine-tuning in SUSY were drawn based on the consideration of a small, limited set of more UV-complete models whose messenger scales never go below $\Lambda=300$~TeV. Whereas in this paper, we consider a broad range of messenger scales down to $\Lambda=20$~TeV, with no attempt at model building. Since we find that only models with $\Lambda<100$~TeV can be better than 10\% fine-tuned after ICHEP, there is in fact no contradiction with the work of \cite{Arvanitaki:2013yja}.}

Throughout this work, we will neglect the role of the gravitino, and in particular the possibility that the higgsino LSP decays to it within the detector.  This is compatible with the low mediation scale $\Lambda\lesssim 100$~TeV provided that the effective SUSY breaking felt by the messengers was much smaller than the ultimate SUSY-breaking scale $\sqrt{F}$ in the hidden sector. (This is the parameter called $k$ in \cite{Giudice:1998bp}.) An assumption along these lines is also necessary to make the baryonic RPV scenario compatible with the low messenger scale; otherwise with very light gravitinos, new proton decay channels such as $p\to\psi_{3/2}K^+$ arise, and proton stability bounds ($\lambda''_{ijk}<(10^{-6}-10^{-15})(m_{3/2}/1$~eV), depending on flavor indices \cite{Choi:1998ak}) would preclude the possibility of hiding SUSY by trading MET for RPV jets.

We conclude by showing rough estimates of the LHC reach for sparticles through 300~fb$^{-1}$ and 3~ab$^{-1}$, using an extrapolation based on the method of \cite{salamprojection}. As will be seen, we are situated at approximately the middle of the rapid rise in the superpartner reach due to the increased LHC energy; after reaching approximately 50 fb$^{-1}$ of data (tentatively expected for Moriond 2017), the LHC will have spent most of its energy boost and additional coverage will be slower and more incremental. Overall, we expect the asymptotic improvement in the LHC reach to be an across-the-board increase of $\sim 900-1200$~GeV to the current limits, largely independent of the SUSY particle mass. 
These projections imply that the high-luminosity LHC (HL-LHC) can exclude or discover all models of fully-natural SUSY that we consider in this work.

These extrapolations assume no  qualitatively new analysis techniques will be developed, so there could be room for even greater future improvement. In particular, the recasted limits we have derived on some of our simplified models come from searches that were generally not designed with the phenomenology of the natural SUSY in mind. Perhaps by targeting natural SUSY (and specifically the simplified models considered here), ATLAS and CMS could significantly improve their sensitivity. For example, the RPV and multi-jet searches  \cite{ATLAS:2016kts,ATLAS:2016kbv,ATLAS:2016uzr,ATLAS:2016kjm,ATLAS:2016nij,ATLAS:2016mnt} are generally optimized for gluino pair production, but we find them to be also relevant for constraining stop pair production. Since the latter involves a different set of physics objects and object multiplicity, with a greatly reduced cross section, perhaps a re-optimization would better maximize $S/\sqrt{B}$ and further extend the reach in this case. For these reasons, we encourage the LHC collaborations to adopt some or all of our simplified models for natural SUSY, for the purpose of optimizing searches and setting official
limits.\footnote{In our work we have conservatively adopted a 50\% ``theory uncertainty'' on our recasting efficiencies, based on our validations against official limit plots. If ATLAS and CMS made official limits on our simplified models, that alone might ``improve'' the limits derived in this work by up to 50\% in the effective cross section.}  The question of naturalness is one of the prime motivations for new physics at the LHC, and as such it is too important to be left solely to amateurs such as ourselves!

Our work suggests many future directions that will be interesting and important to explore. On the collider phenomenology side, given the ability of the LHC searches to powerfully constrain conventional natural SUSY models, as demonstrated here, it will be important to examine 
models with more exotic final states (see \cite{Evans:2015caa} for a recent brief review). Possibilities may include displaced decays (for the signatures we study, these were already well constrained by 8~TeV data \cite{Liu:2015bma,Csaki:2015uza}, while displaced leptons are less tested \cite{Evans:2016zau}); and collimated particles that fail isolation such as ``dirty leptons'' \cite{Graham:2014vya,Brust:2014gia} and ``lepton jets'' \cite{ArkaniHamed:2008qp,Baumgart:2009tn,Cheung:2009su}. Also, we have ignored tau leptons in this work, assuming that they will be at least as stringently constrained by either jets+MET or lepton-based searches.  It might be worthwhile to test this assumption more rigorously, e.g.\ in the case of displaced taus where there are important gaps in coverage \cite{Evans:2016zau}.

On the model building side, there are several, well-known extensions of the MSSM (e.g.~Dirac gauginos) which provide loopholes to the tuning bounds. These are not considered here, but they are increasingly well-motivated. We will discuss them further in Section \ref{sec:conclusion}. 
Even within the more conventional context of natural SUSY with light higgsinos, stops and gluinos, there are many interesting model building directions to pursue. As discussed above, we will show that $\Delta\leq10$ requires a very low messenger scale ($\Lambda\lesssim 100$~TeV), and this is an important constraint of future models of natural SUSY. This is especially true for models of Effective SUSY, since these constructions generally tie SUSY breaking to the generation of flavor in the SM. This would mean that flavor must also be generated at an extremely low scale, and it is not at all obvious that this is viable. Some examples of previous attempts include Refs.~\cite{ArkaniHamed:1997fq,Gabella:2007cp,Aharony:2010ch,Craig:2011yk,Hardy:2013uxa}. Also, we find that adding a HV/Stealth sector to the MSSM can trade MET for jets and greatly reduce the bounds. It is interesting to speculate whether this additional sector could be used for anything else, such as dark matter or raising the Higgs mass. More generally, obtaining a 125~GeV Higgs is a major issue for natural SUSY and requires going beyond the MSSM, and it is interesting to think about whether extensions of the MSSM which succeed in raising the Higgs mass could also help to hide SUSY at the LHC.

This paper is organized as follows. In Section~\ref{sec:recast}, we describe our methodology for reinterpreting LHC searches (with additional information provided in Appendix~\ref{app:recast}). The models of natural SUSY we consider are described in Section~\ref{sec:models}, and the resulting experimental limits on these models in Section~\ref{sec:bounds}. We conclude in Section~\ref{sec:conclusion} with projections for the future reach of the LHC and model-building directions suggested by the existing constraints.

\section{Recasted searches and methodology}
\label{sec:recast}

In the following sections, we will consider the status of natural SUSY after the most recent results from the 13~TeV LHC, as mostly reported in the ICHEP 2016 conference \cite{ICHEP16}. These results, using $12-18$~fb$^{-1}$ of data from CMS and ATLAS, greatly extend the experimental reach of the LHC for gluinos and squarks. 
We concentrate on the searches listed in Table \ref{tab:searches}, each of which has many signal regions (SRs) that target specific mass spectra and supersymmetric production modes.\footnote{We also considered the ATLAS 7-10 jets+MET search \cite{Aad:2016jxj} with 3.2/fb and the CMS black hole search \cite{CMS:2015iwr} with 2.2/fb. Due to the strong possibility of control-region contamination, the latter necessitated a conservative reinterpretation along the lines of \cite{Evans:2013jna}. Neither search set the strongest limit in any of the simplified models considered in this paper, so they are not included here. However, an update of the CMS BH search to the full $\sim 15$~fb$^{-1}$ dataset (especially if optimized for gluino production as in \cite{Khachatryan:2016xim}) would likely have competitive sensitivity to high-multiplicity, non-MET simplified models such as the RPV-based scenarios considered here. }
\begin{table}[t]
\begin{center}
\begin{tabular}{|c|c|c|}\hline
Search                       & Data (fb$^{-1}$) & Reference \\\hline
ATLAS 2-6 jets + MET  &       13.3       & \href{https://cds.cern.ch/record/2206252/}{\tt ATLAS-CONF-2016-078} \cite{ATLAS:2016kts} \\\hline
ATLAS 8-10 jets + MET   &       18.2       &\href{https://cds.cern.ch/record/2212161/}{\tt ATLAS-CONF-2016-095} \cite{ATLAS:2016kbv} \\\hline
ATLAS $b$-jets+MET  & 14.8 & \href{https://cds.cern.ch/record/2206134/}{\tt ATLAS-CONF-2016-052} \cite{ATLAS:2016uzr} \\\hline
CMS jets + MET  ($H_T^{\rm miss}$) &       12.9       & \href{https://cds.cern.ch/record/2205158/}{\tt CMS-PAS-SUS-16-014} \cite{CMS:2016mwj} \\\hline\hline
ATLAS SSL/3L  &  13.2 &\href{https://cds.cern.ch/record/2205745/}{\tt ATLAS-CONF-2016-037} \cite{ATLAS:2016kjm} \\\hline
ATLAS 1L+jets+MET & 14.8 &\href{https://cds.cern.ch/record/2206136/}{\tt ATLAS-CONF-2016-054} \cite{ATLAS:2016lsr} \\\hline\hline
ATLAS multi-jets (RPV) & 14.8 &\href{https://cds.cern.ch/record/2206149/}{\tt ATLAS-CONF-2016-057}  \cite{ATLAS:2016nij} \\\hline
ATLAS lepton+many jets & 14.8 &\href{https://cds.cern.ch/record/2211457/}{\tt ATLAS-CONF-2016-094} \cite{ATLAS:2016mnt} \\\hline
\end{tabular}
\end{center}
\caption{Searches most important to our study. All use the 13~TeV LHC data.}
\label{tab:searches}
\end{table}

As in \cite{Evans:2013jna}, we did not recast searches with photons or two or more opposite-sign leptons, under the assumption that any natural SUSY scenario yielding these signatures would be even more constrained than the simplified models we have considered here.

As can be seen from Table~\ref{tab:searches}, we primarily use ATLAS searches. In most cases, the CMS searches have so many signal regions (100+) that they are difficult to reinterpret. A proper approach would require sophisticated statistical methods combining multiple exclusive bins, using information (the correlation matrix of errors) that is not publicly available. 
In contrast, the ATLAS searches explicitly provide 95\% CL limits on number of events due to new physics for each signal region, and generally have far fewer, coarser bins, allowing us to simply use the most sensitive SR to set a conservative (but reasonably accurate) limit.

One important case where CMS did include aggregate signal region information is the jets+MET search with $H_T^{\rm miss}$ \cite{CMS:2016mwj}, which we find to be very powerful. The CMS jets+MET search has $b$-tagged categories, low-MET and high-MET categories, few jet and many jet categories. As a result, it is equivalent to the union of several different ATLAS jets+MET searches.\footnote{Other CMS general-purpose searches \cite{CMS:2016alpha,CMS:2016xva} provide an equivalent reach on  simplified models studied but provide significantly less information. \cite{CMS:2016alpha} has no aggregate signal regions and does not even provide observed event counts in the text! \cite{CMS:2016xva} does have aggregate signal regions, but adds considerable computational complexity in calculating the clustered jets used in the $M_{T2}$ variable, and was beyond the scope of this work. }

Our simulation methodology is as follows. We generate hard events using \textsc{MadGraph 5.2.3.3} \cite{Alwall:2014hca}, generating pairs of gluinos, squarks, and antisquarks in all possible combinations. Decay and showering is performed via \textsc{Pythia 8.219} \cite{Sjostrand:2014zea} (with which we implement the RPV and stealth decays of the higgsino), and we use \textsc{Delphes 3.3.2} \cite{deFavereau:2013fsa} with an ATLAS-approximating detector geometry for detector simulation.
In order to speed up the event generation, we worked with unmatched samples in most cases.\footnote{Because of the choice of models studied in this paper (in particular, given the high number of jets in the unmatched events), matrix/element/parton-shower matching is usually not necessary: the only case where including extra jets at the generator level makes a difference in the experimental acceptance is for the RPV/Stealth models. We refer to Section~\ref{sec:resultsRPV} for further comments.
}
The NLO cross sections for the superparticle production are obtained via \textsc{Prospino} \cite{Beenakker:1996ed,Beenakker:1996ch,Beenakker:1997ut}. The cutflows for the ATLAS and CMS analyses are recasted using \textsc{Root 5.34} \cite{root}, and validated against the published experimental limits on supersymmetric simplified models. 

The details of the simulated analyses, along with the results of these validation checks, are shown in Appendix~\ref{app:recast}. As seen there, while our simulation technique is largely successful in matching the experimental results, sometimes there are slight differences due  to the crudeness of our recasting framework.
Sources of error could include: the use of \textsc{Delphes} instead of a complete \textsc{Geant4}~\cite{Brun:1994aa} detector simulation; the use of the ATLAS detector geometry for CMS searches; or the use of the 70\% $b$-tagging efficiency working point for all searches, while some of the experimental searches use the 77\% working point. (This latter choice significantly increases the $c$-quark mistagging rate: thus, the use of lower efficiency working point will not always reduce our limits when compared to the official searches, especially if the search involves $b$-vetoes.) Another possible issue is that each of our validation plots (based on a simplified SUSY model) has limits set by a small number of signal regions, while our general models might be sensitive to different signal regions, which are thus never explicitly validated against the experimental efficiencies. 

In any event, as can be seen in Appendix~\ref{app:recast}, the addition of  a 50\% ``recasting uncertainty'' (i.e.~multiplying or dividing the signal efficiencies by a factor of 1.5) around the baseline results is sufficient to bring our recast in line with the official limits, in almost every case. For the models of natural SUSY analyzed below, we will show exclusion limits in which we have taken the lower end of our recasting uncertainty. That is,  we consider a point in the mass parameter space excluded if it exceeds the observed limit from any of the SRs for each search in Table~\ref{tab:searches}, after we have lowered our efficiencies by a factor of 1.5. We use this conservative estimate to make sure that we do not falsely exclude parameter space because of possible inaccuracies in our recasting procedure. 

\section{Overview of simplified models} \label{sec:models}

Here, we will give a brief overview of the various simplified models that we will use in the next section to illustrate the current status of natural SUSY. Although our simplified models are based on the MSSM for the most part (the HV/Stealth scenario is one exception), we expect they are representative of a much broader class of natural SUSY models, and therefore our qualitative conclusions should be much more general. 

In all the models we consider, the lightest supersymmetric particle in the MSSM spectrum is the higgsino, which for reasons of naturalness should be lighter than 300~GeV. (We do not consider alternative models where the higgsino mass arises not from $\mu$ but from some SUSY-breaking operator \cite{Dimopoulos:2014aua,Martin:2015eca,Cohen:2015ala,Garcia:2015sfa,Delgado:2016vib}. We will discuss this possibility further in Section \ref{sec:modelbuilding}.) At present, the best limits on direct higgsino production are set by LEP, with $\mu \geq 95$~GeV for a stable LSP \cite{Heister:2002mn,Abbiendi:2002vz} and $\mu\geq103$~GeV for RPV decays of the higgsino into $udd$ quarks \cite{Costantini:2002jw,Barate:2000en}. (As discussed in the Introduction, we assume that the gravitino is such that the higgsino does not decay to it within the detector.) The recent CMS search with opposite-sign leptons and MET \cite{CMS:2016zvj} does not yet set limits on direct higgsino production, but bodes well for the future in finally overtaking LEP limits. We will take two benchmark values of the higgsino mass, either $\mu=300$~GeV (the maximum allowing $\Delta=10$) or $\mu=100$~GeV (at the LEP limit).

One might wonder if a stop or gluino LSP  (possibly NLSP with a gravitino LSP) is allowed in any part of the parameter space: if the colored partner is (meta-)stable, searches for  $R$-hadrons \cite{CMS:2015kdx} set limits at 1.6 and 1~TeV for gluinos and stops, respectively. For a colored NLSP with prompt decays to a gravitino LSP, the usual simplified topologies with a massless neutralino apply (e.g.~$\tilde{t}\to t\psi_{3/2}$) and both stops and gluinos limits are well above 300~GeV. If the decays violate $R$-parity, stop and gluino LSPs are excluded by pair-produced multi-jet resonances, up to 400 GeV \cite{Khachatryan:2014lpa,ATLAS:2016sfd,CMS:2016pkl} and 800 GeV \cite{Chatrchyan:2013gia,Aad:2015lea}, respectively. 
 Similarly, the presence of additional light neutralinos and charginos (as might be expected alongside the higgsino LSP) will only add to the decay chains of colored spartners, with more final states more easily picked up by the searches considered here. Therefore, we do not believe that we have introduced significant blind spots by assuming a natural higgsino LSP.

For each choice of $\mu$, we consider the following models: 
\begin{enumerate}
\item {\it ``Vanilla SUSY''}: the MSSM with $R$-parity conservation (RPC) and all three generations of squarks degenerate.\footnote{Note that we set both left- and right-handed squarks at the same scale. A variant of this model would have right-handed down-type squarks decoupled: this does not affect the fine-tuning (which is affected by both chiralities of stops, but not by the right-handed sbottom) but reduces the SUSY cross sections by removing the production of $\tilde d_R, \tilde s_R$. In this case, the limits on squarks are reduced by about 100 GeV with respect to the Vanilla SUSY model (shown in Fig.~\ref{fig:resultsstable}).
}
\item {\it ``Effective SUSY''}: The RPC MSSM with all squarks other than the stops and the left-handed sbottom decoupled from the mass spectrum (see Refs.~\cite{Dimopoulos:1995mi,Cohen:1996vb}). For the light third-generation fields, their (pole) masses are taken to be the same, but our results will still hold if they are of the same order.

\begin{figure}[t]
\includegraphics[width=0.85\columnwidth]{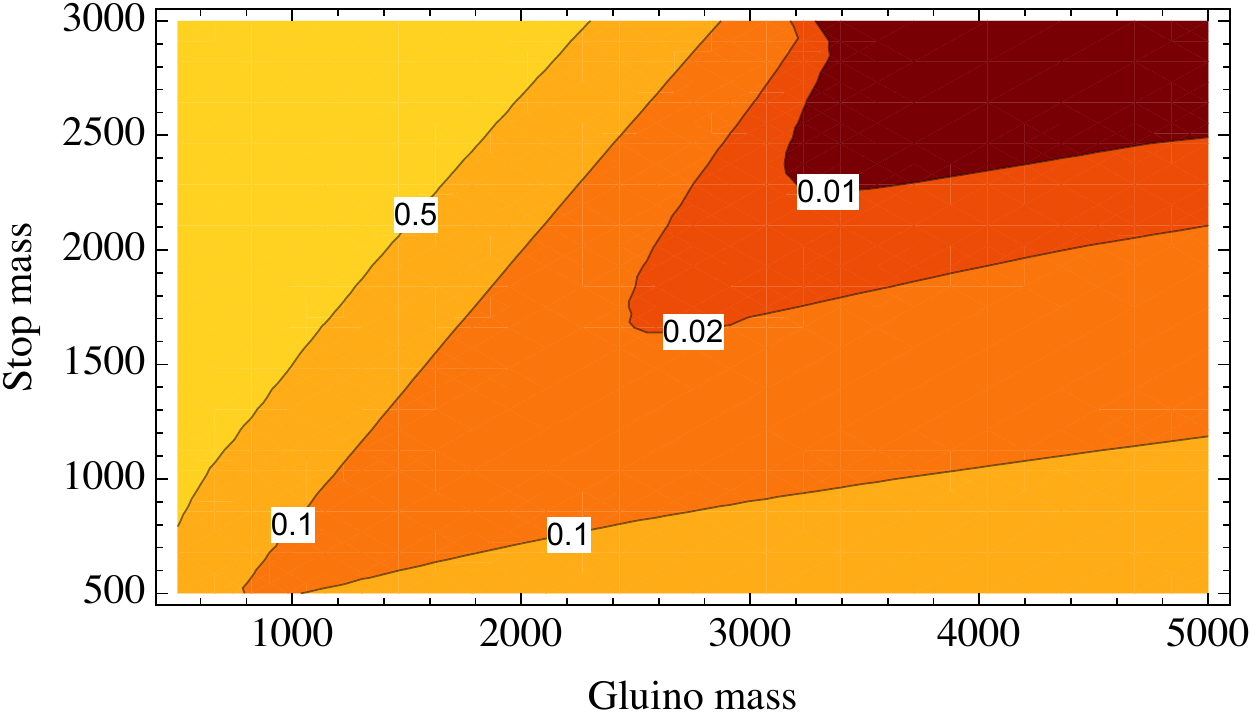}
\caption{Ratio of the 13~TeV total cross section for gluino, squark, and antisquark pair production (in all combinations) in the Effective SUSY model compared to the vanilla SUSY cross section, as a function of gluino and stop masses. \label{fig:effective_xs}}
\end{figure}

The key advantage this model has over the vanilla MSSM in evading the LHC constraints is the much reduced cross section for colored SUSY sparticle production, as the 1st and 2nd generation squarks are heavy. With a Majorana gluino, the $t$-channel gluino-mediated valence-squark cross sections are enormous, assuming the squarks are kinematically accessible. In Fig.~\ref{fig:effective_xs}, we show the ratio of the total colored SUSY production cross section in the Effective SUSY scenario over the vanilla SUSY case. In the region of most interest for natural SUSY, when both the gluinos and squarks masses are $\sim 1-2$~TeV, the cross section for colored pair production is reduced by more than a factor of 10. Obviously, this significantly reduces the experimental reach for these models.

As we briefly discussed in the Introduction (see \cite{Buckley:2016tbs} for details), decoupled 1st/2nd generation squarks tighten (loosen) the naturalness bound on the stops (gluino), because they lower the stop IR masses through the two-loop RGEs (raise the gluino masses through the one-loop pole mass corrections).  In the ``Effective SUSY'' models we consider here, we set 1st/2nd generation squarks to what amounts to a ``sweet spot'' at 5 TeV: too heavy to be efficiently produced at the LHC, heavy enough to help with the gluino fine tuning, but light enough that they do not reduce the natural stop mass range by much.

\item {\it RPV SUSY}: 
The MSSM with baryonic $R$-parity violation (RPV) in which the lightest higgsino can itself decay promptly to three quarks (see \cite{Barbier:2004ez} for a review). By turning the MET from the higgsino LSP into more jets via RPV decays, the SUSY signal can be hidden (to some degree) in the larger QCD background.  Although there are many possible flavor combinations for $UDD$ RPV, we will focus on the 
$cds$ operator as the one which is relatively safe from precision constraints \cite{Barbier:2004ez},
yet will relax experimental limits the most. We note that the limits from the will largely carry over to other flavor combinations as well (except for top quarks).\footnote{Baryonic RPV scenarios in which the $td_id_j$ operator dominates (e.g.~motivated by flavor symmetries \cite{Nikolidakis:2007fc,Csaki2012a,Monteux:2013mna}) result in top-rich final states which either give large-radius jets or leptons. We have checked this case is better constrained:  the limits are raised by about 200~GeV for both gluinos and squarks/stops.
}

\item {\it HV/Stealth SUSY}: an R-parity conserving ``hidden valley'' (HV) extension of the MSSM \cite{Strassler:2006qa,Strassler:2006im} in which the higgsino can further decay into new gauge-singlet scalar $S$ and its fermionic superpartner $\tilde S$:
\[
\tilde{H} \to S \tilde{S},~ S \to gg,
\]
The scalar $S$ decays with 100\% branching ratio into pairs of gluons, while the fermion is stable due to RPC. This model trades MET for jets, provided that  $m_S+m_{\tilde S}\approx m_{\tilde H}$ and $m_{\tilde S}\approx 0$. (We take  $m_S=m_{\tilde H}-10$~GeV, and $m_{\tilde{S}}=5$~GeV.) This simplified model is also a proxy for a number of different well-motivated scenarios. For instance, it could also arise from GMSB with higgsino NLSP,  $\tilde H\to h+\tilde G$, with $m_{\tilde H}\approx m_h$ \cite{Meade:2009qv}. It can also be thought of as a ``lite'' version of Stealth SUSY \cite{Fan:2011yu,Fan:2012jf}; embedding our particle spectrum into an actual Stealth SUSY construction would provide a natural explanation of the required mass degeneracy, while presumably not modifying the limits significantly. 

\item {\it Effective SUSY with RPV}: a combined Effective-RPV SUSY scenario with  first and second generation squarks decoupled and a higgsino LSP decaying to three jets via baryonic RPV. Not surprisingly, this scenario is the least constrained by current searches.
\end{enumerate}
In the models with an unstable higgsino, we assume the conventional cascade decays until reaching the higgsino LSP: for example, the chargino (nearly-mass degenerate with the neutral higgsino) decays to the LSP via an off-shell $W$ and is not allowed to decay directly to jets.

\section{Results}
 \label{sec:bounds}
 
Finally, we are ready to explore the implications of the latest LHC null results for natural SUSY.  We will use our recasting framework to calculate the limits on the simplified models described in the previous section, and then overlay the $\Delta\le 10$ ``fully natural'' region (as determined using the precision calculations  in \cite{Buckley:2016tbs}) over these limits to see what range of natural gluino and stop masses are still allowed. As discussed in the Introduction and described in detail in \cite{Buckley:2016tbs}, the fully-natural $\Delta\le 10$ region is wedge shaped, because the physical stop mass now depends sensitively on the gluino mass through the RGEs, while the physical gluino mass depends to a lesser extent on the stop masses through finite thresholds. For other choices of $\Delta$, the limits on the masses scale approximately  as $\sqrt \Delta$. The extent of these regions depends on the messenger scale $\Lambda$, with lower $\Lambda$ leading to larger allowed masses, see  the LL formulas  (\ref{deltamhusqhiggsino})-(\ref{deltamhusqgluino}) for the rough, qualitative intuition. In our plots, we show the tuning wedges for $\Lambda = 20$ and 100~TeV.\footnote{While $\Lambda=10^{16}$~GeV is well-motivated,  the $\Delta\le 10$ tuning wedge for this case is so tiny that it would barely show up in the plots, so we do not bother to show it.
In fact, for $\Lambda=10^{16}$~GeV, the $\Delta\le 100$ tuning region happens to be qualitatively similar to the $\Lambda=20$~TeV, $\Delta\le 10$ region, see \cite{Buckley:2016tbs} for details.}
 
Our results from applying the reinterpreted LHC searches to the natural SUSY models, scanning over gluino and stop masses, are summarized in Fig.~\ref{fig:resultsstable} for vanilla SUSY, Fig.~\ref{fig:resultsEff}, for Effective SUSY, Fig.~\ref{fig:resultsRPVHV} for both RPV SUSY and HV/Stealth SUSY, and Fig.~\ref{fig:results3rdRPV} for Effective RPV SUSY. Of the eight available experimental analyses in Table~\ref{tab:searches}, we only show the most constraining for each model, to avoid cluttering the plots. A dashed line indicates the combined nominal limits that we find in absence of any recasting uncertainty; solid lines indicate the limits with the aforementioned conservative reduction in signal efficiency by a factor of 1.5. For reference, we also include the appropriate limit on each model from the 8~TeV data, using the recasting framework developed in Refs.~\cite{Kats:2011qh,Evans:2012bf,Evans:2013jna}.

\subsection{Vanilla SUSY}

\begin{figure}[h!]
\includegraphics[width=0.85\columnwidth]{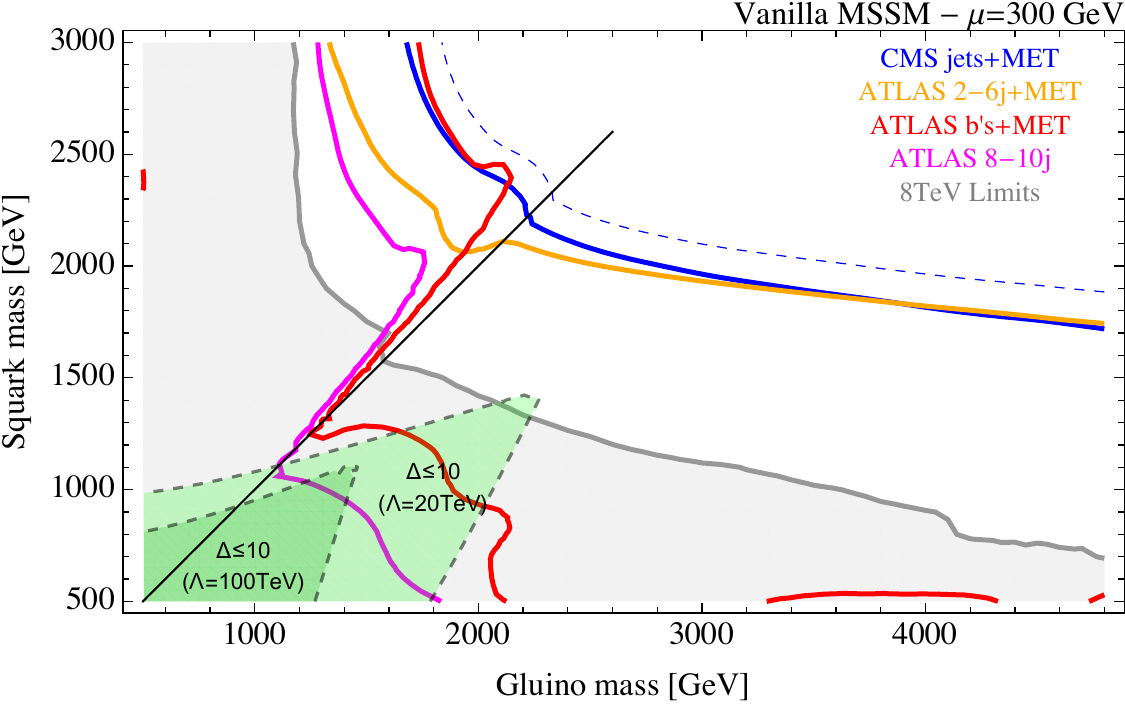}
\caption{
Limits on the ``vanilla'' SUSY model with  
$m_{\tilde{H}}= \mu=300~\gev$  as a function of the gluino and the degenerate squarks masses. The combined limits for $\mu=100~\gev$ are virtually indistinguishable and we do not show them.
All limits are conservative as they already include a factor of 1.5 efficiency reduction to account for possible recasting errors. The nominal limits without recasting uncertainty are shown in dashed blue. The gray shaded area was already excluded by 8 TeV searches (using the framework of Refs.~\cite{Kats:2011qh,Evans:2012bf,Evans:2013jna}). The shaded green regions with dashed lines show the $\Delta \leq 10$ naturalness bound on the gluino and stop masses for  $\Lambda = 20$ and 100~TeV.
\label{fig:resultsstable}}
\end{figure}

 In Fig.~\ref{fig:resultsstable} we see that the parameter space for natural vanilla SUSY (i.e.~the MSSM with light higgsinos, gluinos, and flavor-degenerate squarks) with $\Delta \le 10$ is completely excluded by the LHC results, even for $\Lambda = 20$~TeV. In fact, this was basically true even at 8 TeV. 
Not surprisingly, the most powerful analyses for constraining this scenario are the general-purpose jets+MET searches. The combination of the large production cross section of gluinos and squarks and the large missing momentum carried away by the LSP due to the short decay chain makes vanilla SUSY a rather easy target for these searches. 
 
 Though not shown in Fig.~\ref{fig:resultsstable}, we find the current 13 TeV limits correspond to $\Delta=20$ (i.e.\ 5\% tuning) with a low messenger scale, or $\Delta=200$ with $\Lambda=10^{16}$~GeV (this is discussed further in Section~\ref{sec:resultssummary}). Increasing $\Lambda$ only reduces the natural region for the gluino and squark masses, as expected from  the LL formulas  (\ref{deltamhusqhiggsino})-(\ref{deltamhusqgluino}). 
 
 Some of the limits exhibit a sharp discontinuity along $m_{\tilde g}=m_{\tilde q}$, becoming much weaker below the diagonal. Above the diagonal, where $m_{\tilde q}>m_{\tilde g}$, the gluino decays dominantly to Higgsinos via off-shell stops and sbottoms, so the signal is top and bottom rich. Below the diagonal, where $m_{\tilde q}<m_{\tilde g}$, the gluino decays to all flavors of squarks with nearly equal branching ratio. This reduces the jet multiplicity and the number of $b$-jets on average, significantly weakening the limits from the searches that require these signatures.

\subsection{Effective SUSY}

\begin{figure}[ht]
\includegraphics[width=0.48\columnwidth]{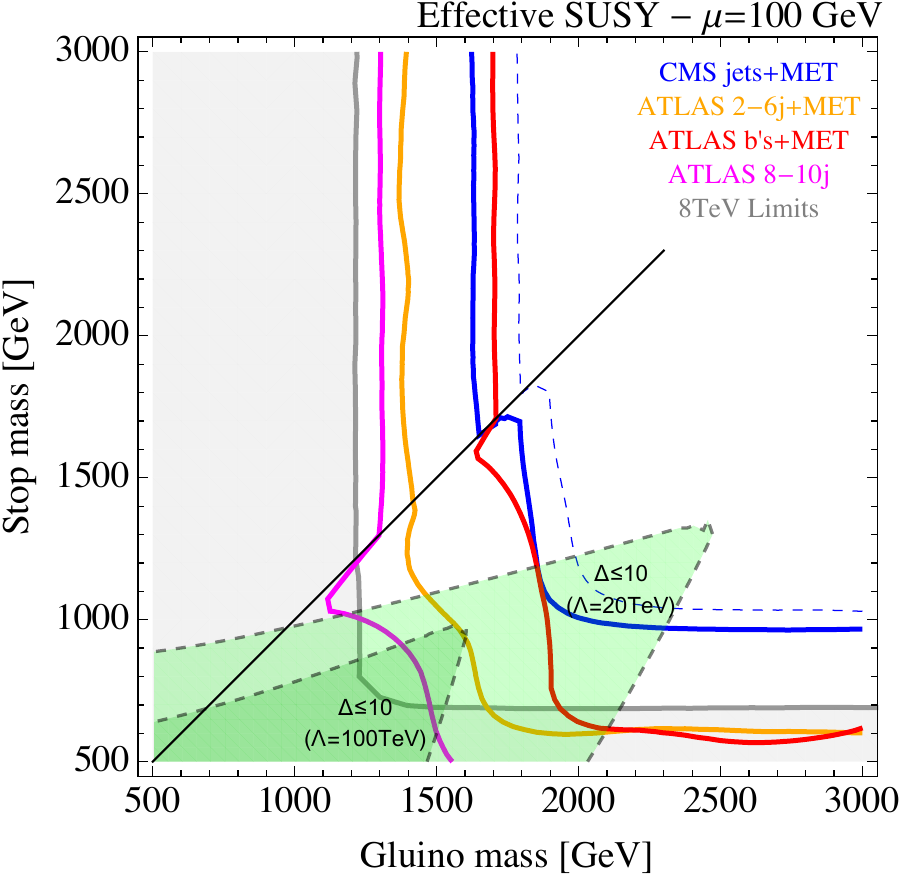}\includegraphics[width=0.48\columnwidth]{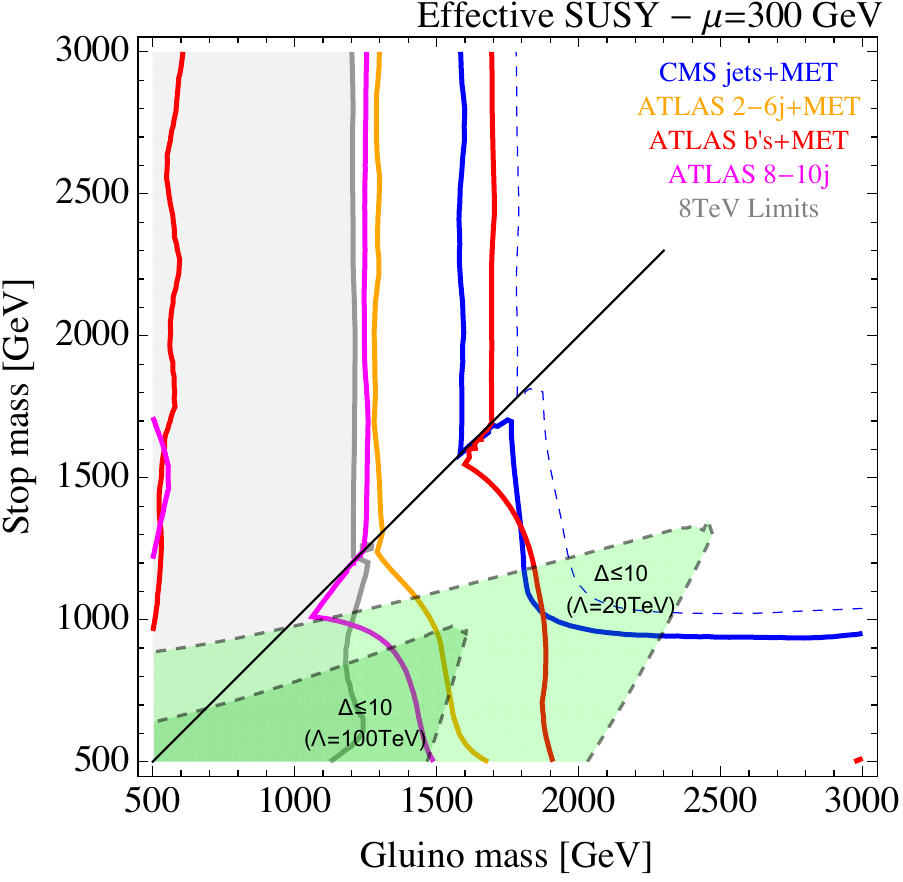}
\caption{Limits on the Effective SUSY model with $\mu=100~\gev$ (left) and 300 GeV (right) as a function of the gluino and the stops and left-handed sbottom masses. The masses of the first and second generation squarks are set to 5~TeV. All other conventions are as in Fig.~\ref{fig:resultsstable}.
\label{fig:resultsEff}}
\end{figure}

We show the limits on the Effective SUSY model (with 1st/2nd generation squarks decoupled to 5~TeV) in Fig.~\ref{fig:resultsEff}: gluinos below 1.8 TeV are excluded while limits on direct stop production are at 900 GeV, significantly raising the previous 8 TeV limits.

Again, we see that the strongest constraints are set by searches targeting large MET, in this case the ATLAS $b$'s+MET and the CMS jets+MET searches. While we did not reinterpret  the many dedicated stop searches from ATLAS and CMS for this study, as can be seen, the general SUSY searches are very powerful, excluding stops nearly up to 1~TeV. Indeed, from the CMS official summary plot \cite{CMSICHEP2016},
one sees that the general purpose CMS jets+MET is nearly as effective as the dedicated stop searches in constraining the basic $\tilde t\to t+\chi_1^0$ simplified model. So we expect that including the dedicated stop searches would not qualitatively change the conclusions here.

Despite these strong limits, there remains a viable (albeit small) range of natural gluino and stop masses in Effective SUSY, but only for extremely low values of $\Lambda$. While $\Lambda =20$~TeV is not yet ruled out, $\Lambda = 100$~TeV is already excluded. Evidently, reducing the SUSY cross section by a factor of $\sim 10$ (see Fig.~\ref{fig:effective_xs}) by decoupling the first and second generation squarks is not enough to completely relax the constraints from the latest round of searches.

\begin{figure}[h!]
\includegraphics[width=0.48\columnwidth]{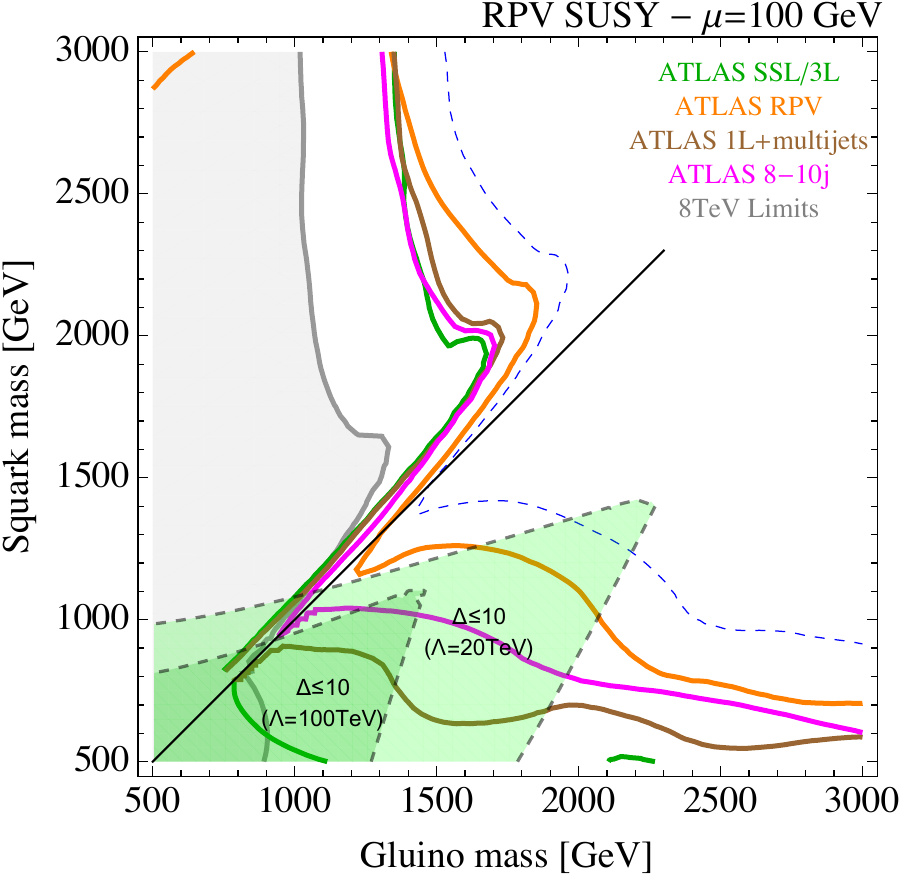}
\includegraphics[width=0.48\columnwidth]{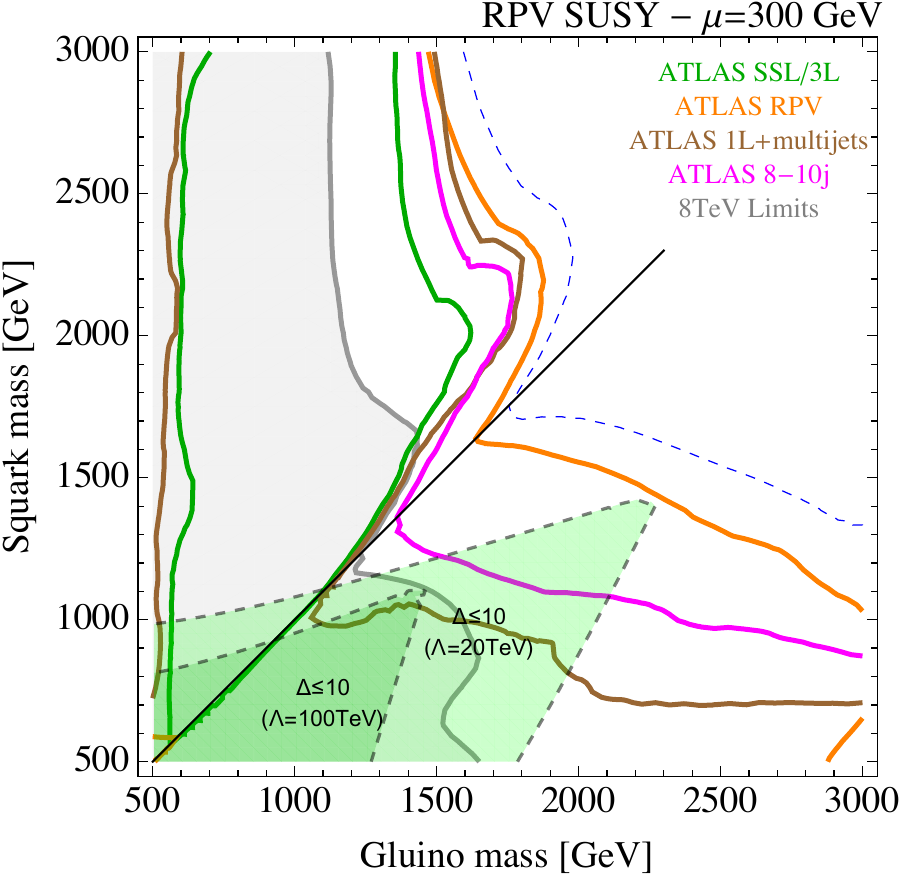}
\\\vskip0.2cm
\includegraphics[width=0.48\columnwidth]{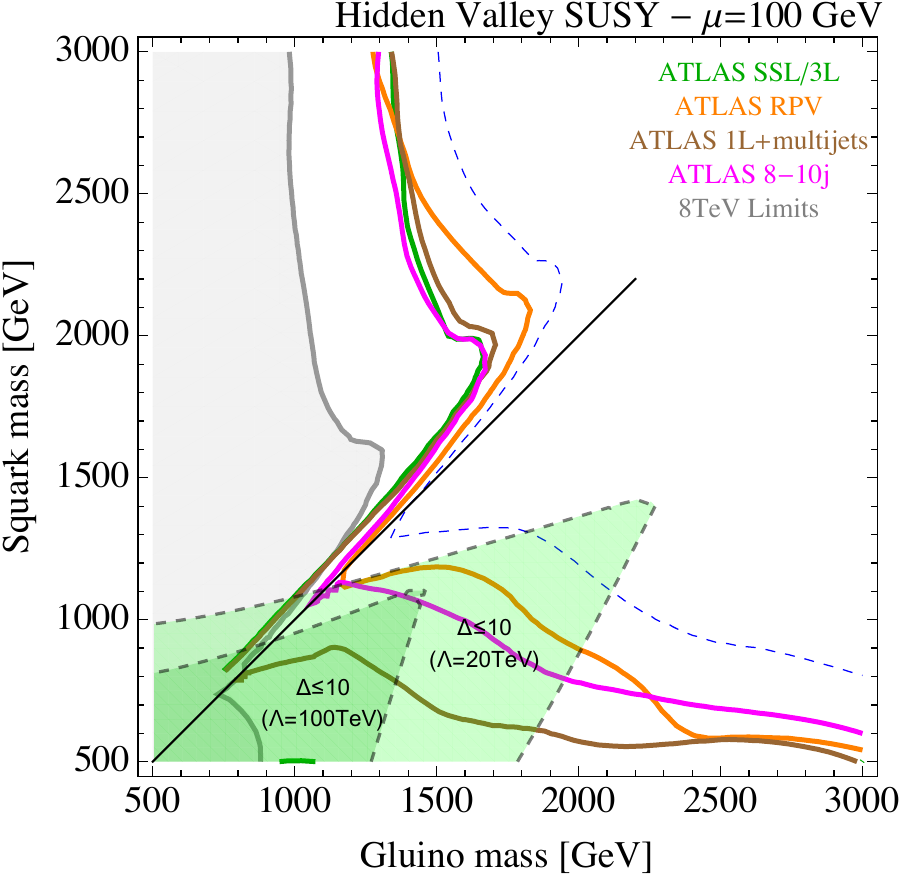}
\includegraphics[width=0.48\columnwidth]{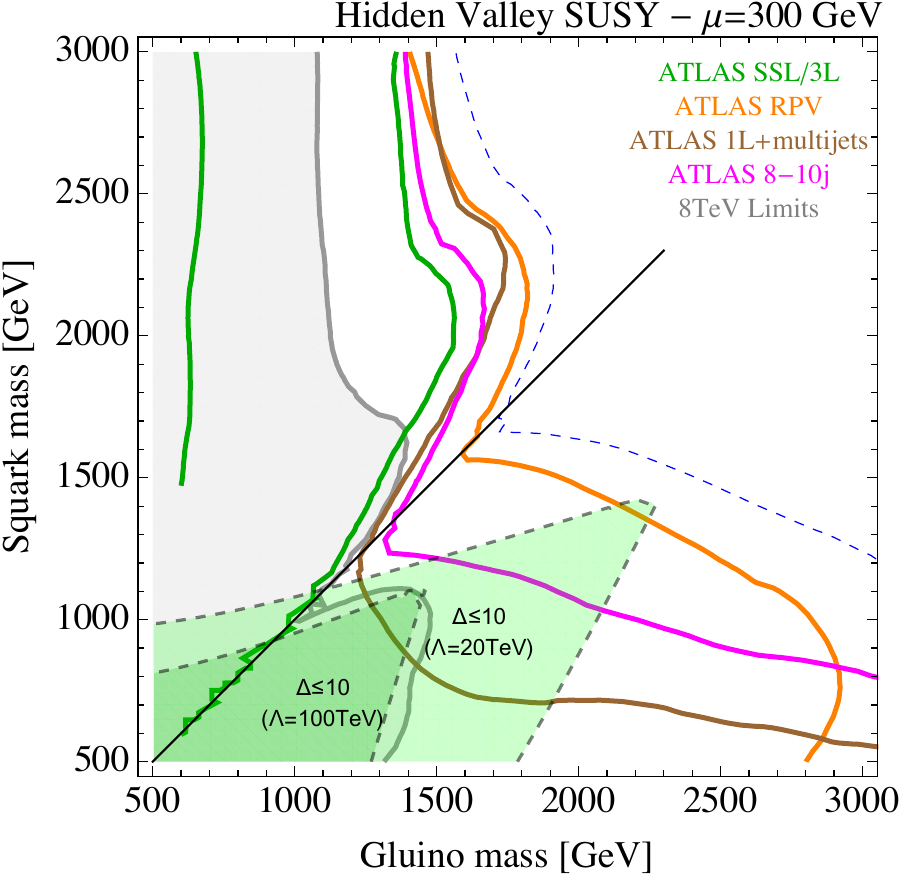}
\caption{Limits on the RPV (top) and HV/Stealth SUSY (bottom) models with $\mu=100~\gev$ (left) and 300 GeV (right) as a function of the gluino and the degenerate squarks masses. All other conventions are as in Fig.~\ref{fig:resultsstable}.
\label{fig:resultsRPVHV}}
\end{figure}

\subsection{RPV and HV/Stealth SUSY}\label{sec:resultsRPV}

We now turn to SUSY models which trade MET for jets. Obviously, these models are going to be far less constrained by the standard MET-based searches. However, searches which target large multiplicities of high-$p_T$ jets instead of MET (such as the ATLAS 8-10 jets search \cite{ATLAS:2016kbv} and the ATLAS RPV search \cite{ATLAS:2016nij}), are still very powerful. In these scenarios, we have included one additional jet at the generator level (and matched the matrix-element and parton-shower calculations in the MLM scheme \cite{Mangano:2006rw,Alwall:2007fs,Alwall:2008qv}): for squarks, the hard process would have resulted in 8 final partons, and adding an extra parton raises the reach of the ATLAS RPV and ATLAS 8-10 jets searches by approximately 100~GeV.

In Fig~\ref{fig:resultsRPVHV}, top row, we show the limits on RPV SUSY, allowing the higgsino to decay into a trio of $cds$ quarks (the results would be similar for any $u_{i\neq3}d_jd_k$ operator, in particular, for final states with $b$ quarks the ATLAS RPV limits would increase, while ATLAS 8-10 jets would stay the same as it does not involve b-tagging).
As can be seen, while the natural masses are excluded for $\Lambda = 100$~TeV, a small region of the $\Lambda = 20$~TeV gluino and squark mass range remains unexplored, assuming the higgsino mass is 100~GeV. If this mass is raised to 300~GeV, the jets resulting from the RPV decay are more effectively captured by the high-multiplicity searches, and the entire $\Delta < 10$ space is excluded. 
With a lighter higgsino, the quarks in the final states are more collimated and result in significantly fewer resolved jets, which is more difficult to distinguish from the QCD background. This important characteristic was discussed at length in  \cite{Evans:2013jna}.

Again, as in the vanilla SUSY case, there is a pronounced shift in the strength of the limits across the $m_{\tilde g}=m_{\tilde q}$ diagonal, because the gluino goes from dominantly decaying to Higgsinos via off-shell stops and sbottoms (above), to decaying to all flavors of squarks equally (below). Above the diagonal, where the gluino decays to Higgsino are top-rich, the ATLAS SS dilepton search \cite{ATLAS:2016kjm} sets an equally strong limit as the high-multiplicity searches.\footnote{This may come as a slight surprise, as the SS dilepton searches were found to be not as effective for constraining natural SUSY at 8 TeV \cite{Evans:2013jna}. The difference is that in the new search, there is a new signal region (SR1b-GG) that does not require any MET. This further highlights the power and importance of doing SUSY searches with low or no MET.} 

The same features are also seen in the HV/Stealth SUSY results, bottom row of Fig.~\ref{fig:resultsRPVHV}. As in the RPV case, $\Lambda = 100$~TeV is already ruled out for HV/Stealth, and only lower values of the messenger scale remain viable. The fact that the limits on the RPV and the HV/Stealth scenarios are so quantitatively similar, despite the scenarios having different kinematics and different number of jets in the final state, is evidence that the LHC limits are fairly robust, and that the simplified models we have chosen are representative of a broader class of scenarios that trade MET for jets. 

\subsection{RPV Effective SUSY}

\begin{figure}[t!]
\includegraphics[width=0.48\columnwidth]{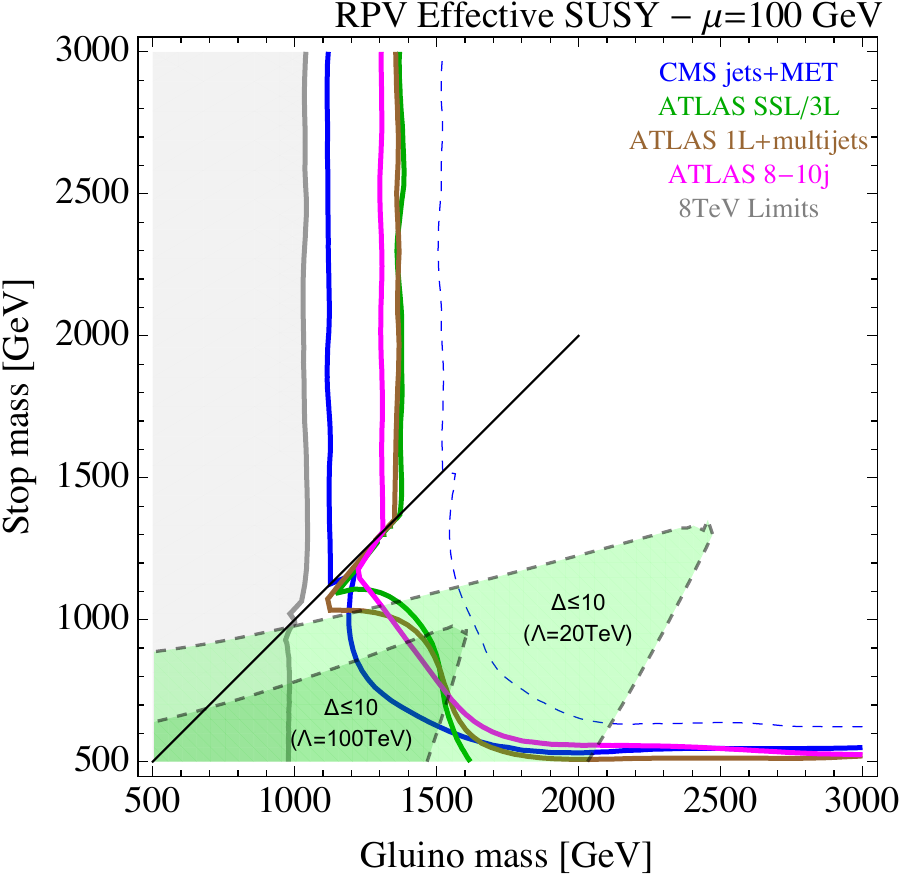}\includegraphics[width=0.48\columnwidth]{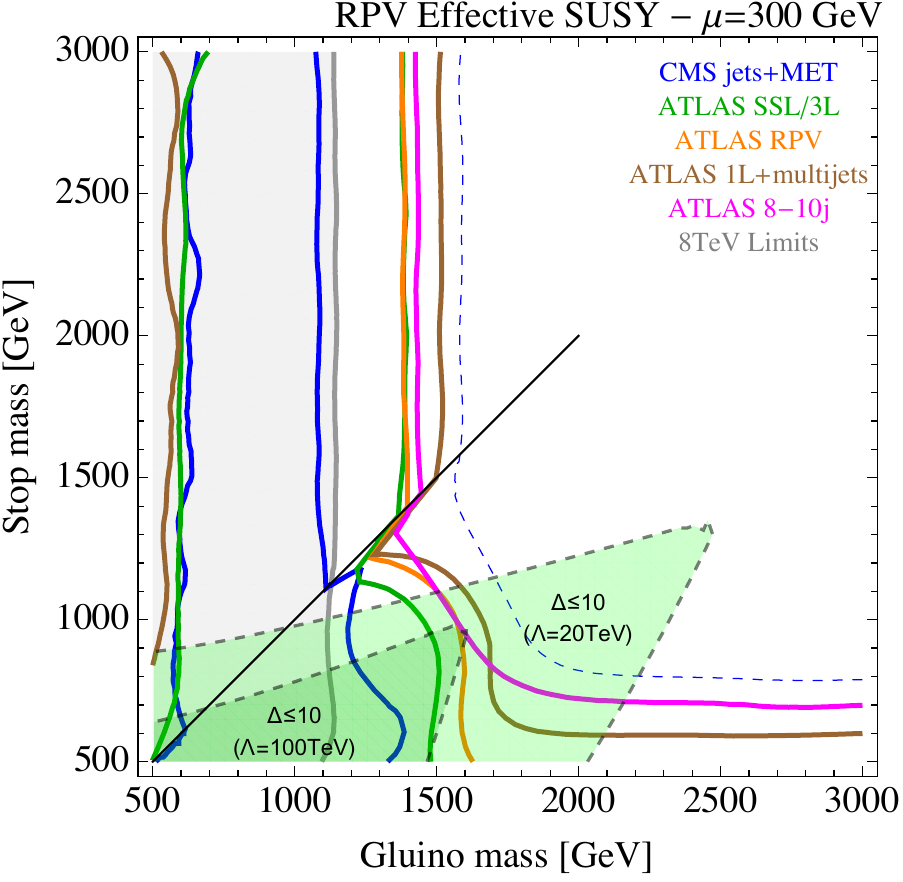}
\caption{Limits on the Effective RPV SUSY model with $\mu=100~\gev$ (left) and $300~\gev$ (right) as a function of the gluino and the stops and left-handed sbottom masses. All other conventions are as in Fig.~\ref{fig:resultsstable}.
\label{fig:results3rdRPV}}
\end{figure}

Finally, having seen the (limited) success of both RPV/HV scenarios and Effective SUSY in evading the LHC bounds on natural SUSY, we consider their combination. In Fig.~\ref{fig:results3rdRPV}, we show the experimental reach for models of Effective RPV SUSY, where the higgsino is unstable and only the two stop squarks and the left-handed sbottom are light, while the remaining sbottom and the first and second generation squarks are decoupled at 5~TeV. As expected, the limits are the weakest of all the models considered so far, with limits on gluinos and third-generation squarks respectively  at 1.4--1.5~TeV and 600--800~GeV. (As in the previous subsection, the same limits would apply to any $u_{i\neq3}d_jd_k$ final state.) Even here, the $\Delta\leq10$ parameter space for a 100~TeV messenger scale is nearly excluded, but much of the $\Lambda = 20$~TeV parameter space remains viable.

\subsection{Summary of results and further implications}
\label{sec:resultssummary}

\begin{figure}[t!]
\includegraphics[width=0.48\columnwidth]{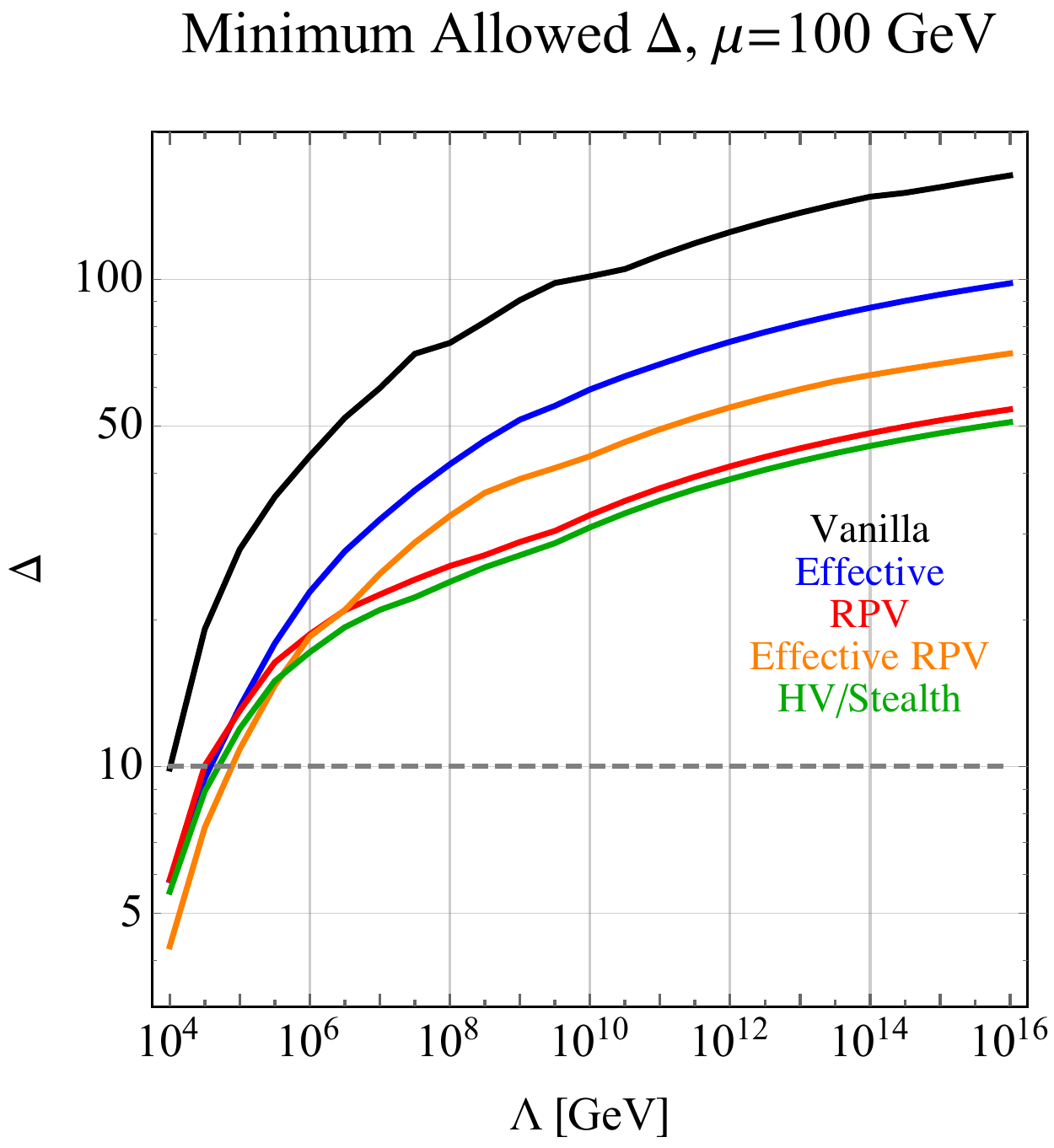}\includegraphics[width=0.48\columnwidth]{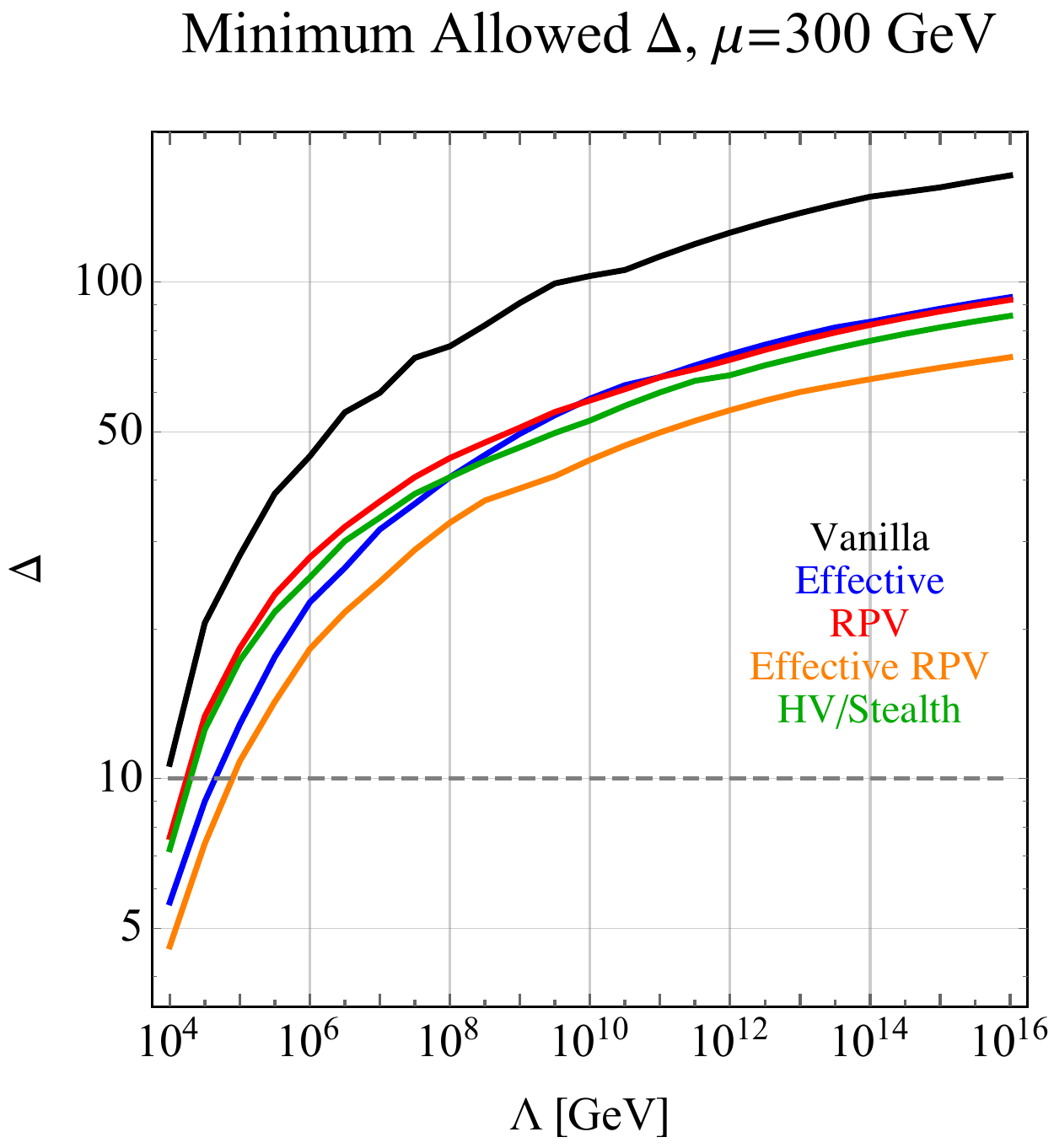}
\caption{Minimum amount of tuning $\Delta$ which is experimentally allowed as a function of SUSY breaking scale $\Lambda$. \label{fig:lambda_tuning} }
\end{figure}

In the previous sections we have excluded a wide range of gluino, squark and stop masses for a variety of natural SUSY models, and understood the implications for fine-tuning. In Fig.~\ref{fig:lambda_tuning}, we further apply the calculations of \cite{Buckley:2016tbs} in order to show
 the minimum amount of tuning $\Delta$ compatible with a given messenger scale $\Lambda$, for each of the natural SUSY models we consider in this paper. (Qualitatively, these curves can be understood/extrapolated from the results shown in the previous subsections, using the LL formulas (\ref{deltamhusqhiggsino})-(\ref{deltamhusqgluino}).) 
 As can be seen, even with our most optimistic scenario (Effective SUSY with RPV decay of higgsinos), the scale $\Lambda$ must be less than 100~TeV for $\Delta\leq10$. It should be noted that other choices of ``acceptable'' levels of fine-tuning allow higher messenger scales. For example every scenario we have considered (except perhaps vanilla SUSY) is only tuned at the percent-level or better, even with messengers at the GUT scale.

Aside from naturalness considerations, the individual recasted limits on each superpartner are noteworthy as they cannot always be obtained from the ATLAS and CMS summary plots (this is particularly true for the RPV/HV/Stealth cases, where the ATLAS RPV and 8-10 jets searches do not consider squark simplified models). For this reason, in Table~\ref{tab:results} we summarize the asymptotic limits on each colored superpartner (gluinos and either mass degenerate squarks or third-generation squarks); these limits are obtained from the plots above by decoupling either the gluinos or the squarks.\footnote{We note that, for mass-degenerate squarks, ``decoupling'' the gluino in practice means taking $m_{\tilde g}\gtrsim10$~TeV, otherwise the gluino $t$-channel contribution to valence squark production remains non-negligible.}

For the models considered so far, we have restricted ourselves to $\mu\leq300$~GeV, due to the tree-level higgsino contribution to fine-tuning. An obvious question is how the limits scale with higher higgsino masses. We find that most of the limits do not change significantly for higgsino masses below $\sim 500$~GeV, but above $\sim 500$~GeV, the MET-based scenarios such as vanilla SUSY and effective SUSY start to see some degradation due to compression, with third generation squarks affected first (near 450 GeV), then degenerate squarks (above 600 GeV), and finally gluinos (above 800 GeV). On the other hand, for models with unstable neutralinos, the limits actually {\it increase} moderately (by $\sim 100-200$~GeV) with higher higgsino masses, as the jets neutralino decays are less boosted. These features are fully consistent with the neutralino mass dependence of the   validation plots shown in the Appendix.


Before we conclude, a comment is required: in both RPV and HV/Stealth models, it can be seen that fully hadronic MET-based searches set limits that are competitive with searches which do not rely on missing energy. A close inspection at individual events reveals that the source of missing energy is neutrinos in the presence of semi-leptonic $W$ decays (for example, from top quarks or $B$ mesons), where the concurring lepton is not isolated and is therefore removed. The resulting events have non-zero missing energy and no isolated leptons, and are accepted by the all-hadronic searches (in particular, CMS jets and ATLAS 8-10 jets, which have moderate MET requirements). While we have closely mirrored the overlap removal procedures in the experimental papers, our cruder simulation framework might be overestimating the reach of these searches. In any case, the more robust limits from other searches (particularly ATLAS RPV and 1L+multijets) in Figs.~\ref{fig:resultsRPVHV} and \ref{fig:results3rdRPV} result in similar exclusions, with differences of at most 50~GeV.

\begin{table}[t]
\resizebox{\columnwidth}{!}{
\begin{tabular}{|c|c|c|c|c|c|c|c|c|c|c|c|} \hline
Model & {Vanilla SUSY} & \multicolumn{2}{c|}{Effective SUSY} & \multicolumn{2}{c|}{RPV SUSY} & \multicolumn{2}{c|}{Stealth/HV SUSY} & \multicolumn{2}{c|}{RPV Effective SUSY}
\\\hline
$\mu$ [GeV] & 100 -- 300 &100  & 300 &100 &300 &100 &300 &100 &300
\\\hline
$m_{\tilde{g}}$ [GeV] & 1730 & \ 1690\, & 1690 & 1310 & 1500 & \ \ 1330 \ \,  & 1440 & \ \ 1350\ \, & 1490
\\
$m_{\tilde{q},\tilde{t}}$ [GeV] & 1500 & 975 & 950 & 700 & 810 & 600 & 750 & 550 & 750
\\\hline
\end{tabular}
\caption{Observed lower limits on the mass of the gluino $m_{\tilde{g}}$ and squarks $m_{\tilde{q}},m_{\tilde{t}}$ for each model considered above. The limits are asymptotic in the sense that they refer to the case where all other superpartners decouple. The vanilla SUSY limits are independent on the higgsino mass for $\mu<300$ GeV.
 \label{tab:results}}
}
\end{table}

\section{Conclusions and Future Directions}
 \label{sec:conclusion}

\subsection{Projections to 300~fb$^{-1}$ and 3~ab$^{-1}$}

We have found in the previous section that while vanilla SUSY has not been fully natural already since Run I, many flavors of SUSY beyond vanilla are currently still viable with $\Delta\le 10$. Here, we show naive extrapolations of our limits to 300~fb$^{-1}$ and 3~ab$^{-1}$ of integrated luminosity (assuming either 13 and 14~TeV collisions), corresponding approximately to the end of the regular LHC operations and the ultimate end of the HL-LHC runs, respectively. 

Our methodology is the same as \cite{salamprojection} (we thank A.~Weiler for some clarifications in this regard). We assume that the signal cross section is controlled by a single mass scale $m$, and that
signal {\it efficiencies} and background {\it counts} remain constant as the integrated luminosity ${\mathcal L}$ and/or center-of-mass (CM) energy $\sqrt{s}$  are increased. The latter makes sense, provided that the mass reach grows in a way that one can always cut harder to keep the backgrounds low, while preserving the sensitivity to new physics at higher and higher mass scales. It also assumes that as ${\mathcal L}$ and/or $\sqrt{s}$ are increased, there are no qualitatively new obstacles that cannot be overcome with more clever analysis techniques. For example, our projections ignore the effect of pile-up: while a challenge for the HL-LHC, it should prove less of a barrier to searches in the mass range above 1~TeV, where SUSY decays typically result in many high-$p_T$ jets and/or large MET.

Using this approach, if the previous limit at CM energy $\sqrt{s_1}$ and integrated luminosity ${\mathcal L_1}$ was at $m_1$, the extrapolated limit to CM energy $\sqrt{s_2}$ and integrated luminosity ${\mathcal L_2}$ can be obtained by  requiring
\beq
m_1^{-2} f(m_1/\sqrt{s_1}) {\mathcal L_1} = m_2^{-2} f(m_2/\sqrt{s_2}) {\mathcal L_2}
\eeq
where $f$ is the parton luminosity (taken here to be $gg$ for simplicity -- the projections for $q\bar q$ initiated production are actually slightly stronger, but qualitatively similar).

\begin{figure}[t!]
\includegraphics[width=0.5\columnwidth]{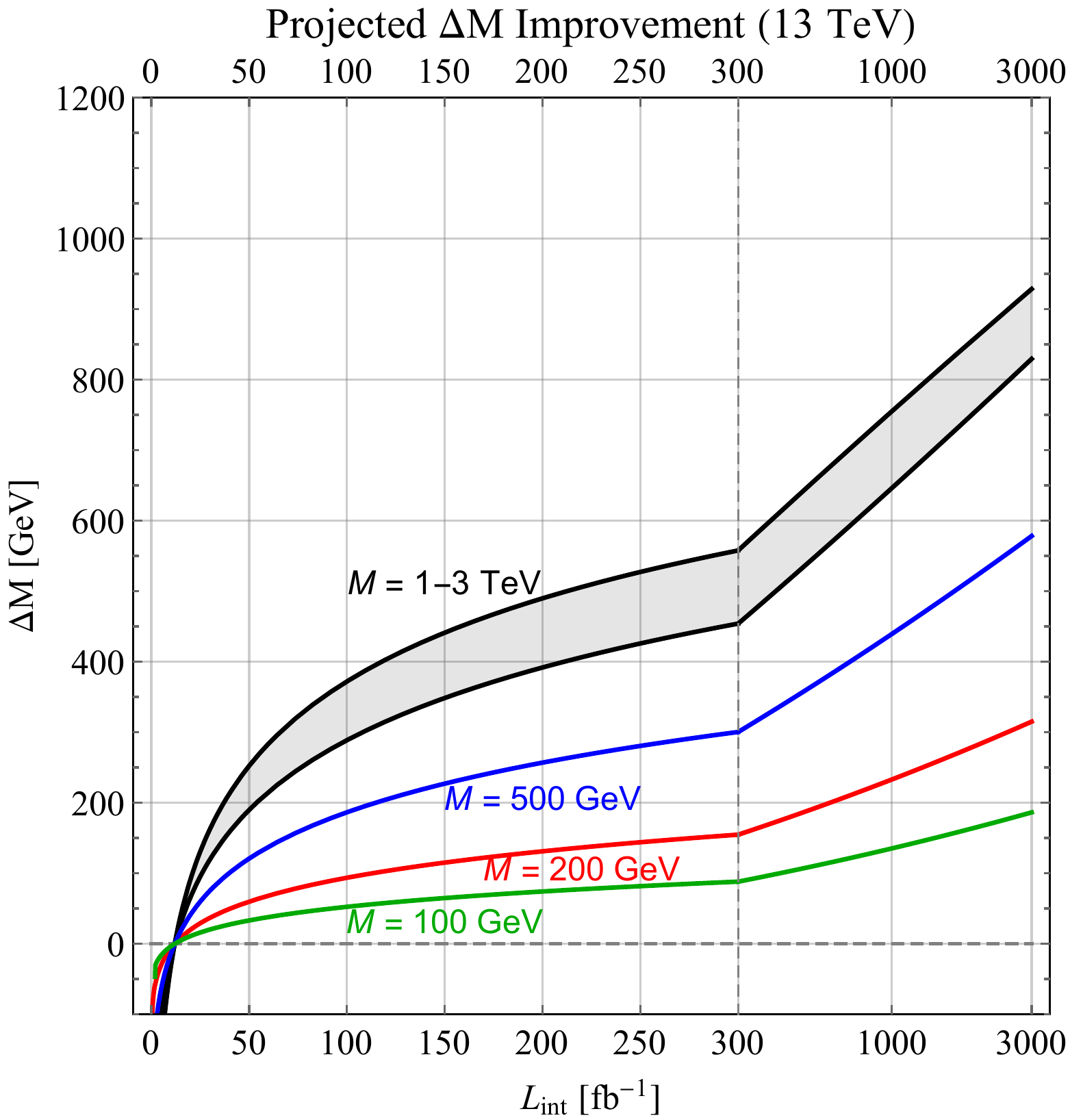}\includegraphics[width=0.5\columnwidth]{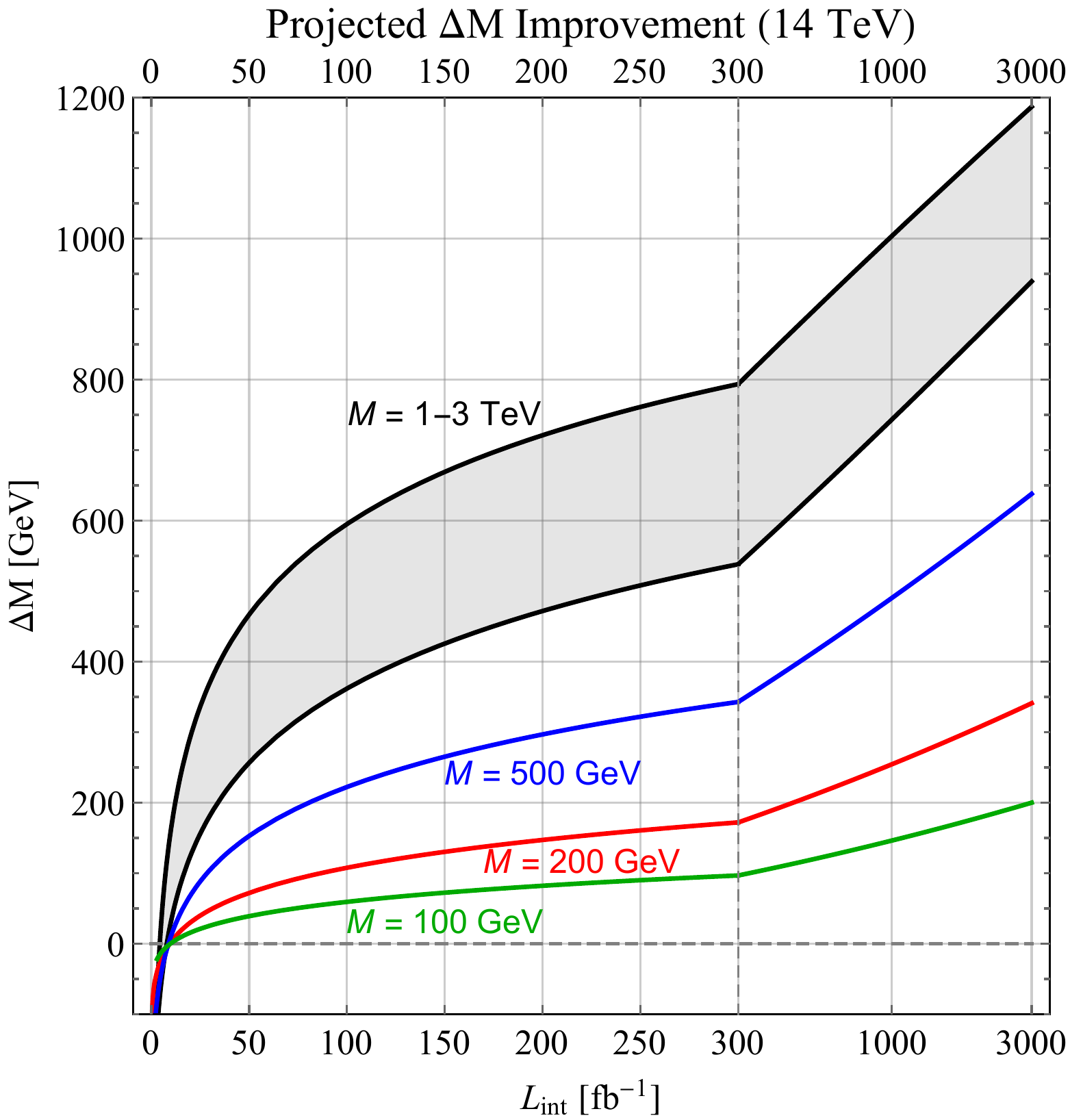}
\caption{Projected improvement $\Delta M$ to the experimental 95\% CL upper limits on a superpartner excluded by 12~fb$^{-1}$ of 13~TeV data, as a function of integrated 13~TeV (left) or 14~TeV data (right). The black shaded region corresponds to the projected $\Delta M$ for an existing limit for a search which currently excludes a particle between 1 and 3 TeV. Here, the improvements are relatively independent of the limit. The green line is the projected improvement for limit which currently excluded a 100~GeV particle, the red line assumes a 200~GeV present limit, and the blue line assumes 500~GeV. The projection technique is as described in the text and \cite{salamprojection}. Note the transition from linear to log scale at 300~fb$^{-1}$. \label{fig:projection}}
\end{figure}

Interestingly, under these assumptions, the ultimate improvement in mass reach, compared to current limits, is nearly a constant shift, $m \to m+\Delta m$, across a wide range of masses  (1~TeV~$\lesssim m \lesssim$~3~TeV). This can be traced back to the fact that, in this range of $x=m/\sqrt{s}$ values, the parton luminosities at the LHC happen to be dropping nearly exponentially, $f(x)\sim e^{-a x}$. In Fig.~\ref{fig:projection}, we show the increase in mass reach benchmarked against a limit set with 12~fb$^{-1}$ of 13~TeV, as a function of integrated luminosity (for both 13 and 14~TeV LHC running). 
After 300~fb$^{-1}$ of data is collected at 13~TeV, the rough rule of thumb is that we expect the limits on both  gluinos and stops to improve by $\Delta M\sim 500$~GeV from their current levels, with a further improvement to $\Delta M \sim 900-1200$~GeV after 3~ab$^{-1}$ at 14 TeV. For example, taking the current best limits on gluino (1.8~TeV) and stop (900~GeV) masses, this would correspond to an ultimate reach of $\approx 3$~TeV and $\approx 2$~TeV respectively.
Using these projections, we see that the $\Delta\le 10$ regions for all models we consider can be fully explored by the end of the HL-LHC era.

As an aside, while very light particles do not see the full shift in mass reach, the asymptotic improvement would still have a very significant impact, fractionally speaking. This is especially relevant for non-colored particles, where the current limits in many cases are not far beyond the LEP bounds. Here an increase of $\sim 200-300$~GeV to the range of experimental sensitivity to these channels would constitute at least a doubling of the existing bounds, if not more. This provides a strong motivation for the construction of the HL-LHC.

\subsection{Future directions for model building}
\label{sec:modelbuilding}

One of the principal values of a recasting work on simplified models is that it motivates further model building to populate the allowed parameter space, which could in turn lead to new correlated signatures to explore (or could suggest that the allowed parameter space cannot be realized in UV complete models). So we will conclude by describing some of the model building avenues that are suggested or motivated by our work.  This is not an exhaustive list, but highlights some directions of theoretical and experimental interest.

The first (and least interesting) possibility in light of these results is simply to relax the requirement of naturalness. The weak scale may involve a slight numerical accident; after all there are plenty of other percent-level accidents in Nature. This is a logical possibility (perhaps further motivated by the 125~GeV Higgs mass), but it may not be experimentally testable. Tuning at the level of $\Delta \sim 100$ is not well constrained at the LHC, and the relevant parameter space is unlikely to be excluded in most SUSY models in the foreseeable future.

The remaining alternatives all require extending SUSY beyond the vanilla scenario. For instance, in this paper we have explored two possibilities. The first, effective SUSY, lowers the total SUSY cross section by removing the possibility of valence squark production. As emphasized in our companion paper \cite{Buckley:2016tbs}, without the addition of further particles as in \cite{Hisano:2000wy}, the squarks cannot be set to arbitrarily high masses without risking tachyonic 3rd generation sparticles. Decoupling some of the squarks also raises questions about the resolution of the supersymmetric flavor problem. All of these difficulties are compounded by the requirement of very low messenger scales ($\Lambda\lesssim 100$~TeV). It remains to be seen whether any viable model of effective SUSY exists with such low messenger scales. Some promising prior work that explored models in this direction includes \cite{ArkaniHamed:1997fq,Gabella:2007cp,Aharony:2010ch,Craig:2011yk,Hardy:2013uxa}.

The second avenue we explored is RPV/HV/Stealth SUSY, which reduces the signal to background ratio by trading MET for jets. It would be interesting to investigate the HV/Stealth sector further, to see whether it could be useful for anything else, e.g.\ dark matter or lifting the Higgs mass. Also, further exploration of RPV scenarios is well-motivated. For instance, a potential option to further reduce the constraints on RPV discussed in this work  might be having at least one large coupling, such that squarks decay dominantly to jets instead of higgsinos (searches for paired dijet resonances are then relevant \cite{Khachatryan:2014lpa,ATLAS:2016sfd}). Although constraints from flavor physics \cite{Barbier:2004ez} can be important, this can be accomodated in models with hierarchies dictated by flavor symmetries \cite{Nikolidakis:2007fc,Csaki2012a,Monteux:2013mna}. At the same time, this also opens the possibility of resonant squark production via the $udd$ operator, with additional signatures and 8~TeV limits recently described in \cite{Monteux:2016gag}. Nevertheless, a low messenger scale calls for cautious model-building if new flavor-violating interactions are to be present.

As noted previously, in this work we have not considered the possibility of decays to light gravitinos inside the detector.  In the high-MET scenarios (vanilla SUSY, effective SUSY), allowing the higgsino to decay to gravitino plus $h/\gamma/Z$ is unlikely to qualitatively change our conclusions on a stable higgsino, with the increase in final state multiplicity possibly reinforcing our limits. One exception would be the compressed scenario $m_{\tilde \chi^0_1}\simeq m_h$, where some of this phenomenology would be covered by our Hidden Valley model, as already noted above. In any event, for what concerns experimental limits, it would be interesting to see if opening a gravitino decay channel would significantly change the limits presented in this work.

Finally, another interesting direction would be to challenge the underlying assumptions going into the tuning calculations: for example the SUSY production cross section can be reduced by introducing ``super-safe'' Dirac gluinos \cite{Fox:2002bu,Kribs:2012gx} (which eliminate valence squark diagrams with $t$-channel gluinos).
Alternatively, models where the higgsino mass is not set primarily by the $\mu$ term, e.g.~\cite{Dimopoulos:2014aua,Martin:2015eca,Cohen:2015ala,Garcia:2015sfa,Delgado:2016vib} are increasingly well-motivated. A higgsino LSP above 600~GeV can lead to compressed spectra and greatly weakened limits, as discussed in Section~\ref{sec:resultssummary}; for instance the gluino could be as light as $\sim 800$~GeV given the current bounds on the $\tilde g\to qq\chi_1^0$ simplified model. Also, if the higgsino was not the LSP, it could lead to weakened bounds if the gluino can decay directly to a HV/Stealth sector \cite{Fan:2015mxp}. 

In this work, we have explored simplified models motivated by natural SUSY where cascade decays of accessible gluinos and stops down to a light higgsino produce collections of ordinary jets, leptons, etc. We have considered scenarios with and without MET. While limits from current searches may still allow fully natural SUSY models, it is a testament to the breadth of the ATLAS and CMS experimental programs that much of the $\Delta\leq10$ parameter space is fully excluded already, while what remains (at $\Lambda\lesssim 100$~TeV) can be completely covered by the end of the HL-LHC run.

\paragraph{Note Added} The first preprint version of this paper differed with respect to the present published version in two aspects: first, it did not include matching, which led to an underestimation of some limits, especially for the ATLAS RPV search constrains on unstable neutralino scenarios of Sec.~\ref{sec:resultsRPV}. Second, a bug was found in Delphes 3.3.2 which caused an overestimation of the ATLAS 8-10 jets search 
limits on the same RPV and Hidden Valley scenarios.\footnote{For more details and the bug-fix appearing in Delphes 3.4.1, see \url{https://cp3.irmp.ucl.ac.be/projects/delphes/ticket/1084}.} The neutralino, which had a significant boost originating from the large mass splitting with the squarks or gluinos, decays to boosted quarks and in the ensuing parton shower a small (but significant enough) fraction of events had long-lived unstable hadrons, with a nominal lifetime large enough to escape the detector. While at the LHC those would deposit all their energy in the hadronic calorimeter and then decay, Delphes would throw away their daughter particles as it used only the list of final stable particles within the detector volume for each event. Given these missing high-momentum particles in some events, the resulting $E_T^{\rm miss}/\sqrt{H_T}$ spectrum was not as steeply falling as it should be, and the requirement $E_T^{\rm  miss}/\sqrt{H_T}>4\gev^{1/2}$ of Ref.~\cite{ATLAS:2016kbv} was satisfied in a larger fraction of events, resulting in stronger limits from that search. Fixing this bug reduced the ATLAS 8-10 jets limits by $200-400$ GeV, especially for the squarks. The net effect of these two changes is negligible in most of the parameter space of Fig.~\ref{fig:resultsRPVHV} with no differences in the combined limits in the natural regions of parameter space and of at most 100 GeV elsewhere.

\section*{Acknowledgements}

We are grateful to C.~Campagnari,  T.~Lari, F.~Moortgat and I.~Vivarelli for their help in clarifying many aspects of the ATLAS and CMS SUSY searches. We also thank N.~Arkani Hamed, N.~Craig, K.~Cranmer, J.~A.~Evans, Y.~Kats, M.~Papucci, M.~Reece, J.~Ruderman, M.~Strassler, S.~Thomas and A.~Weiler for helpful discussions. Finally, we are grateful to J.~A.~Evans, M.~Farina, Y.~Kats, M.~Reece and M.~Strassler for comments on the draft. The work of
AM and DS are supported by DOE grant DE-SC0013678. The work of SM is supported by DOE grant DE-SC0010008.

\appendix

\section{Recasting Details and Validation\label{app:recast}}

Here, we discuss the details of our reinterpretation of each of the ATLAS and CMS experimental searches used in this paper. Each search has provided details of the cut-flow for the various signal regions, along with the expected and observed number of events, allowing us to apply the results to the supersymmetric models of interest. In order to validate our implementation of each search, we apply our recast to the simplified supersymmetric models used by the experiments themselves. These simplified models typically have only two or three supersymmetric particles kinematically accessible. As such, they are not appropriate for our study of naturalness in SUSY.

As in the study of the full supersymmetric theories, described in Section~\ref{sec:recast}, we generate hard events for the simplified models using \textsc{MadGraph5} \cite{Alwall:2014hca}, with NLO cross sections from \textsc{Prospino} \cite{Beenakker:1996ed,Beenakker:1996ch,Beenakker:1997ut}; we have not noticed appreciable differences in the validation plots between generating matched and unmatched samples. We decay and shower these events with \textsc{Pythia8} \cite{Sjostrand:2014zea}. Detector simulation is via \textsc{Delphes3} \cite{deFavereau:2013fsa}, which makes use of the \textsc{FastJet} \cite{Cacciari:2011ma} package for jet finding. We use the same jet-clustering algorithms as in each experimental search, namely the longitudinally invariant $k_t$ algorithm \cite{Catani:1993hr,Ellis:1993tq} and the anti-$k_t$ algorithm \cite{Cacciari:2008gp}, as well as jet trimming \cite{Krohn:2009th} on reclustered jets \cite{Nachman:2014kla}.
In Delphes, we reconstruct the missing energy vector as the negative sum of all calorimeter deposits and all muons. In Delphes a ParticleFlow algorithm is also present, which combines the tracker and the calorimeter information to define physics objects. The differences in the missing energy are usually small, but we choose the former method as it results in validation plots that are in slightly better agreement with the official results, although the differences are within our ``theory error'' estimate.

In most cases, recasting the searches requires only a straight-forward implementation of the cut-flow described in the relevant conference note. Where necessary, we note any deviations we were forced to take from the search as reported by the experimental collaboration. In the following, we show our validation plots, comparing the experimental exclusion region with the exclusion region we find on simplified supersymmetric models. As in the original experimental searches, limits are set by taking observed limits on the number of signal events in the signal region which has the best expected limits.

\subsection{ATLAS Same-sign Lepton/Three Lepton}

The ATLAS note CONF-2016-037 \cite{ATLAS:2016kjm} is a search for supersymmetric particles decaying to jets and leptons, requiring either two same sign leptons or three leptons in the final state. A number of signal regions are defined, separated by number of $b$-tagged jets, lepton multiplicity and missing transverse momentum. 

\begin{figure}[ht]
\includegraphics[width=0.5\columnwidth]{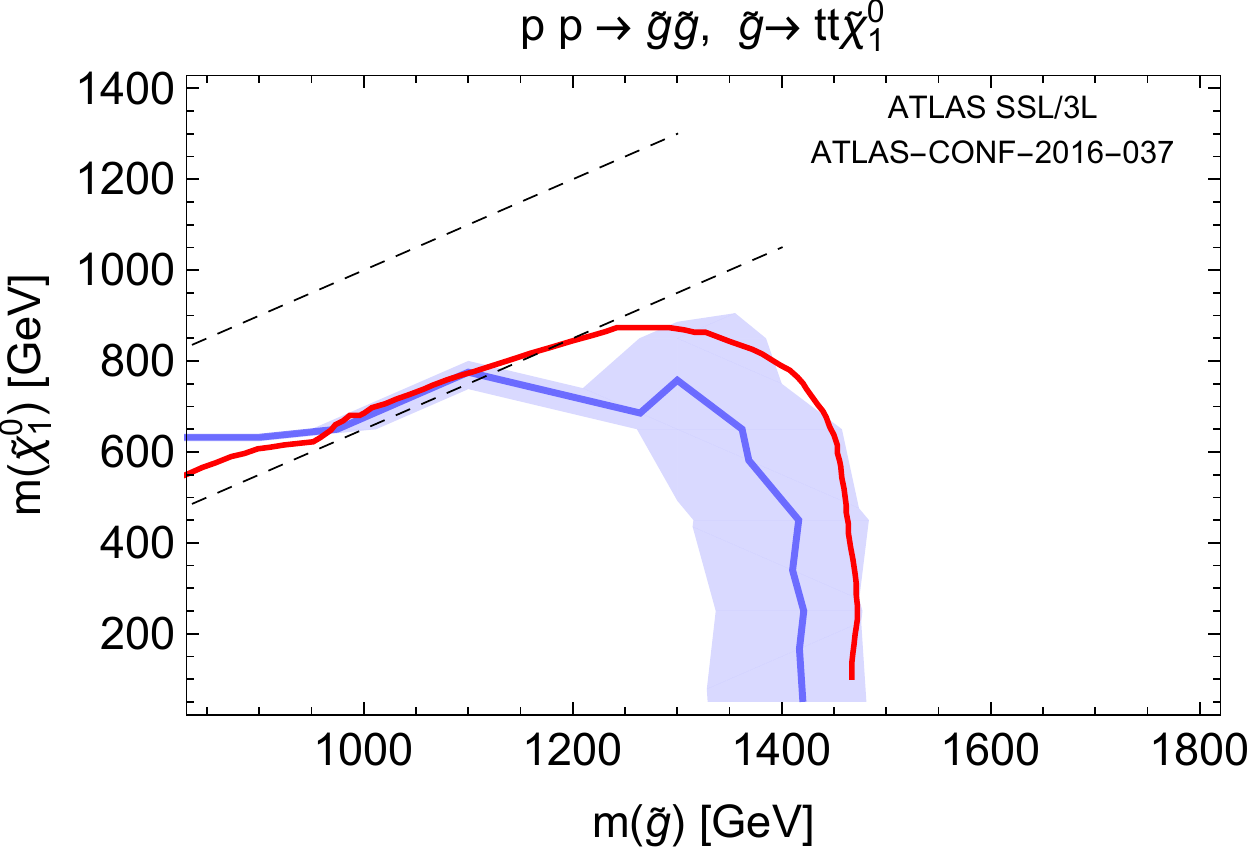}

\caption{Limits on a supersymmetric simplified model from our recasted search of \cite{ATLAS:2016kjm} (blue line) with 50\% error on the number of events in the signal regions (blue shaded region), compared to the experimental results (red line). \label{fig:ATLAS-CONF-2016-037}}
\end{figure}

We validate our recasting of the search by generating events using one of the supersymmetric simplified models considered by \cite{ATLAS:2016kjm}: gluino pair production decaying to top pairs and a neutralino $\tilde{g} \to t\bar{t}\tilde{\chi}_1^0$. The published limits are shown in Fig.~\ref{fig:ATLAS-CONF-2016-037} with the results of our recasted search on simulated data, along with a 50\% ``recasting uncertainty'' on the number of events in each signal region.

\subsection{ATLAS Lepton Plus Jets}

The ATLAS note CONF-2016-054 \cite{ATLAS:2016lsr} is a search for gluinos and squarks decaying through $W^\pm$ bosons (via charginos), requiring one lepton in the final state, along with jets and missing transverse momentum. A number of signal regions are defined, separated by number of jets, $b$-tagged jets, and missing transverse momentum. 

\begin{figure}[ht]
\includegraphics[width=0.5\columnwidth]{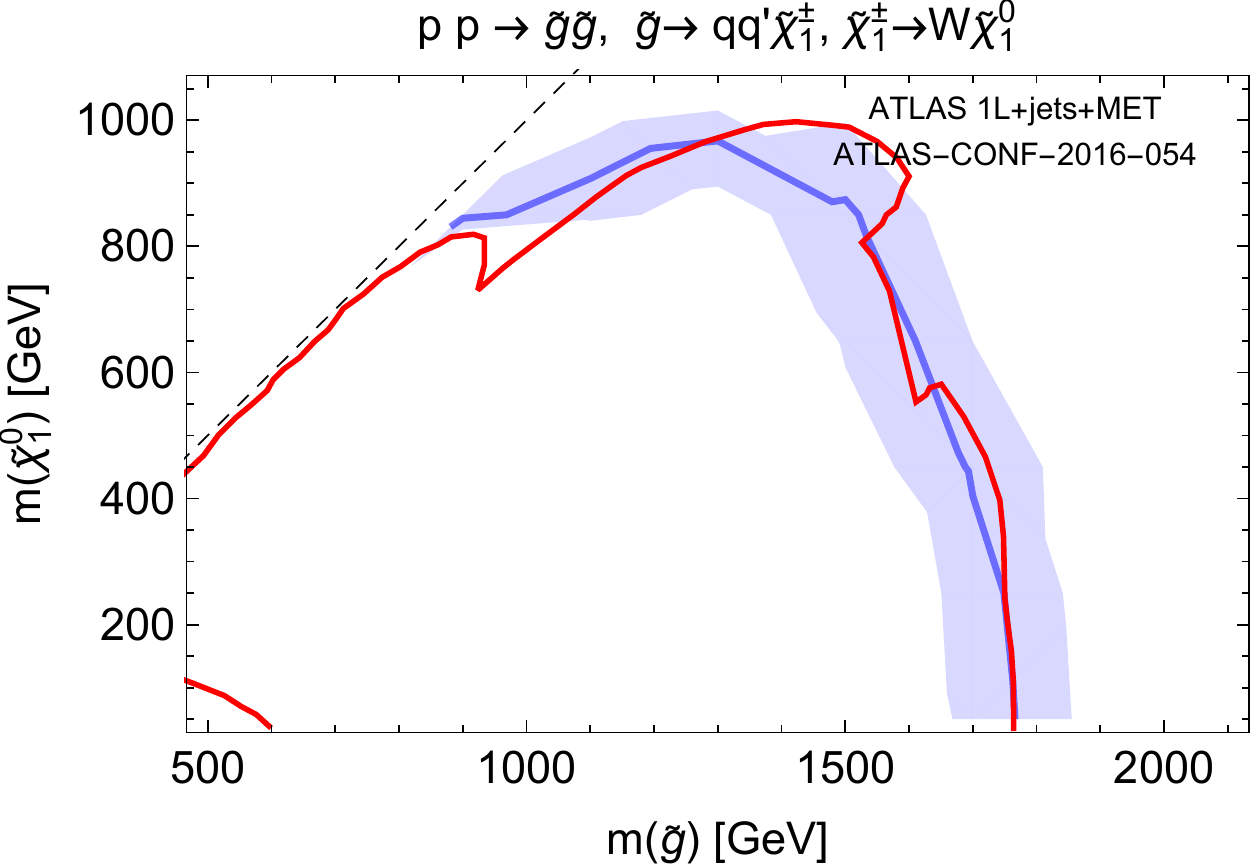}

\caption{Limits on supersymmetric simplified models from our recasted search of \cite{ATLAS:2016lsr} (blue line) with 50\% error on the number of events in the signal regions (blue shaded region), compared to the experimental results (red line). \label{fig:ATLAS-CONF-2016-054}}
\end{figure}

We validate our recasting of the search by generating events using the supersymmetric simplified model considered by \cite{ATLAS:2016lsr}: gluino pair production decaying to light-flavor quarks and a chargino, which itself decays to a neutralino and a $W$ boson: $\tilde{g} \to q\bar{q}'\tilde{\chi}_1^\pm,~\tilde{\chi}^\pm_1 \to W^\pm \tilde{\chi}^0_1$. The chargino mass is set to be the average of the gluino and neutralino mass.  The published limits are shown in Fig.~\ref{fig:ATLAS-CONF-2016-054} with the results of our recasted search on simulated data, along with a 50\% ``recasting uncertainty'' on the number of events in each signal region.

\subsection{ATLAS Multi-$b$}

The ATLAS note CONF-2016-052 \cite{ATLAS:2016uzr} is a search for gluinos decaying to third generation quarks (tops or bottoms) and missing transverse momentum. At least three $b$-jets must be identified in the final state. Some signal regions further require ``fat'' jets which have topological similarities to top quarks. We followed the procedure outlined in \cite{ATLAS:2016uzr} by re-clustering the $\Delta R =0.4$ jets into jets of radius $0.8$ using the anti-$k_T$ algorithm in \textsc{Delphes}, and then further trimming the resulting jets by removing subjets whose $p_T$ falls below $10\%$ of the $p_T $ of the re-clustered jet.

\begin{figure}[ht]
\includegraphics[width=0.5\columnwidth]{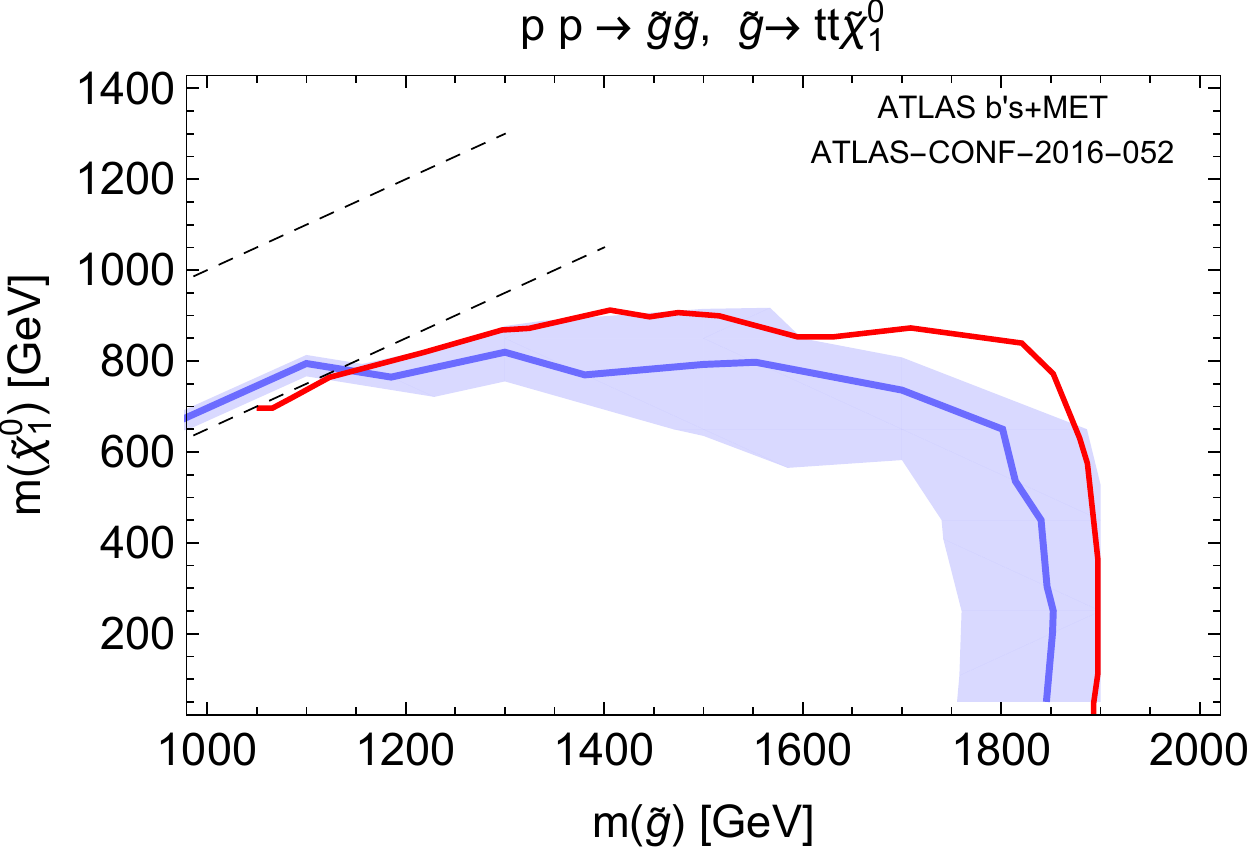}\includegraphics[width=0.5\columnwidth]{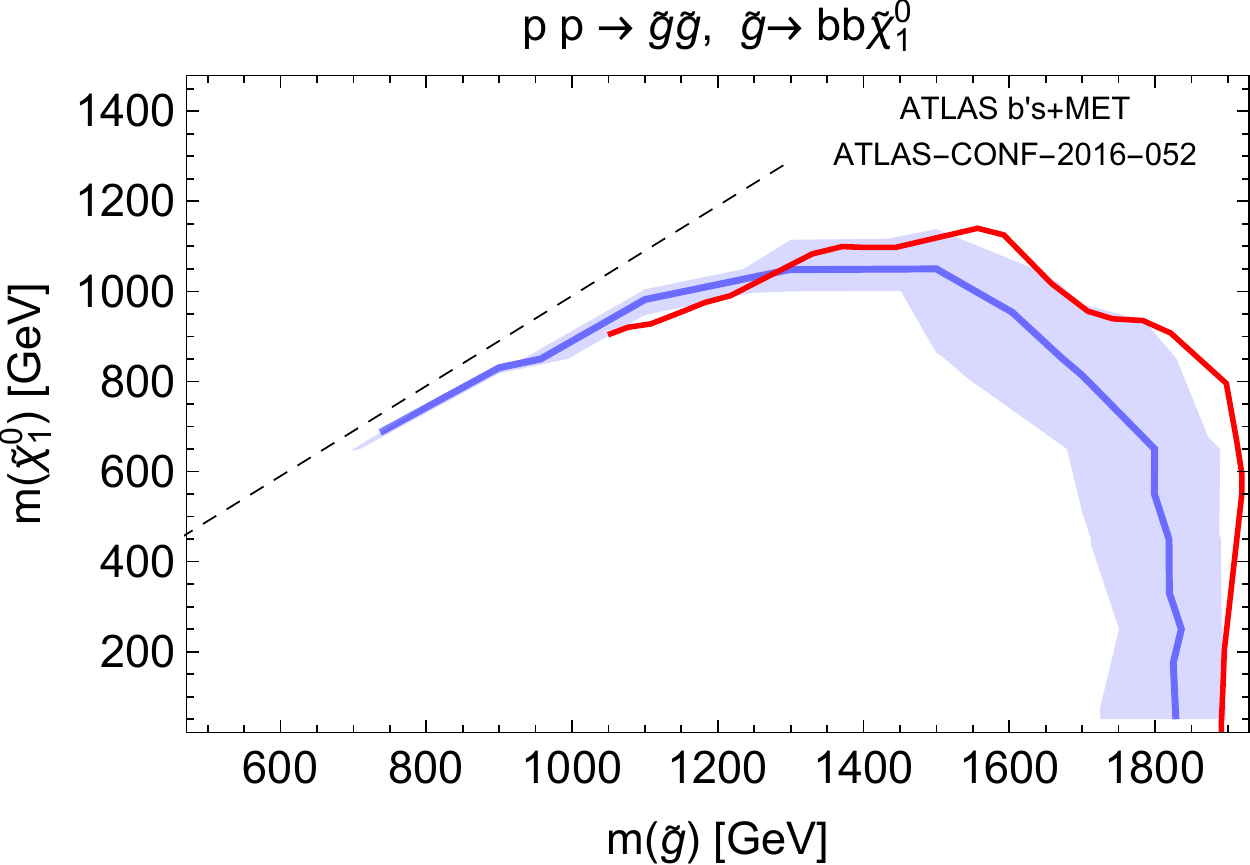}

\caption{Limits on supersymmetric simplified models from our recasted search of \cite{ATLAS:2016uzr} (blue lines) with 50\% error on the number of events in the signal regions (blue shaded regions), compared to the experimental results (red lines). \label{fig:ATLAS-CONF-2016-052}}
\end{figure}

We validate our recasting of the search by generating events using the supersymmetric simplified model considered by \cite{ATLAS:2016uzr}: $(i)$ gluino pair production decaying to top pairs and a neutralino $\tilde{g} \to t\bar{t}\tilde{\chi}_1^0$ and $(ii)$  gluino pair production decaying to bottom pairs and a neutralino $\tilde{g} \to b\bar{b}\tilde{\chi}_1^0$.  The published limits are shown in Fig.~\ref{fig:ATLAS-CONF-2016-052} with the results of our recasted search on simulated data, along with a 50\% ``recasting uncertainty'' on the number of events in each signal region.

\subsection{ATLAS RPV}

The ATLAS note CONF-2016-057 \cite{ATLAS:2016nij} is a search for RPV SUSY. A number of signal regions are identified with varying number of jets, $b$-tagged jets, and large radius jets. These ``fat'' jets are simulated in \textsc{Delphes} by reclustering the calorimeter jets into jets of radius $1.0$ using the anti-$k_T$ algorithm. Then, the resulting large jets are trimmed by re-clustering their components using the $k_T$ algorithm with a sub-jet radius parameter of $0.2$ and discarding sub-jets carrying less than $5\%$ of the original large jet. The surviving sub-jets are used to calculate the ``fat'' jet energy and momentum, which is then further corrected by the jet energy scale.

\begin{figure}[ht]
\includegraphics[width=0.5\columnwidth]{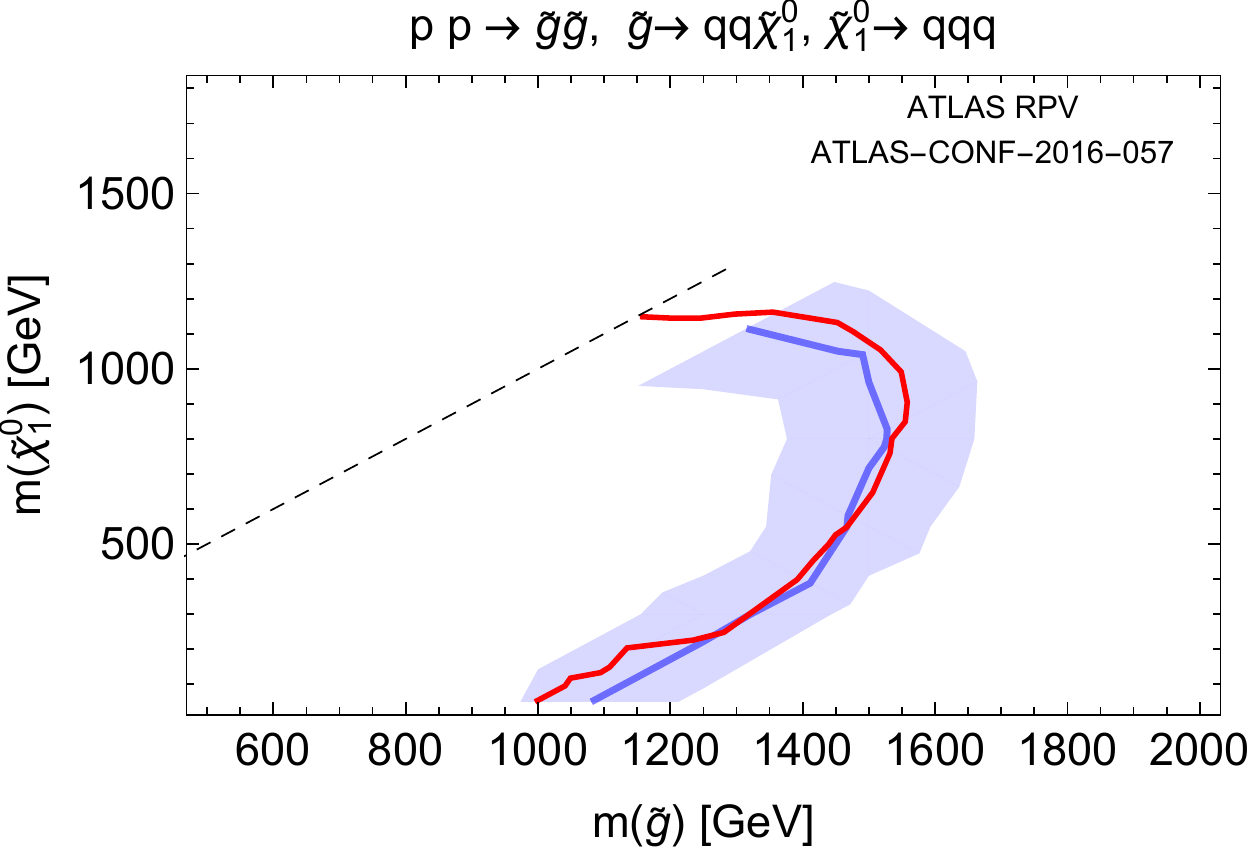}\includegraphics[width=0.5\columnwidth]{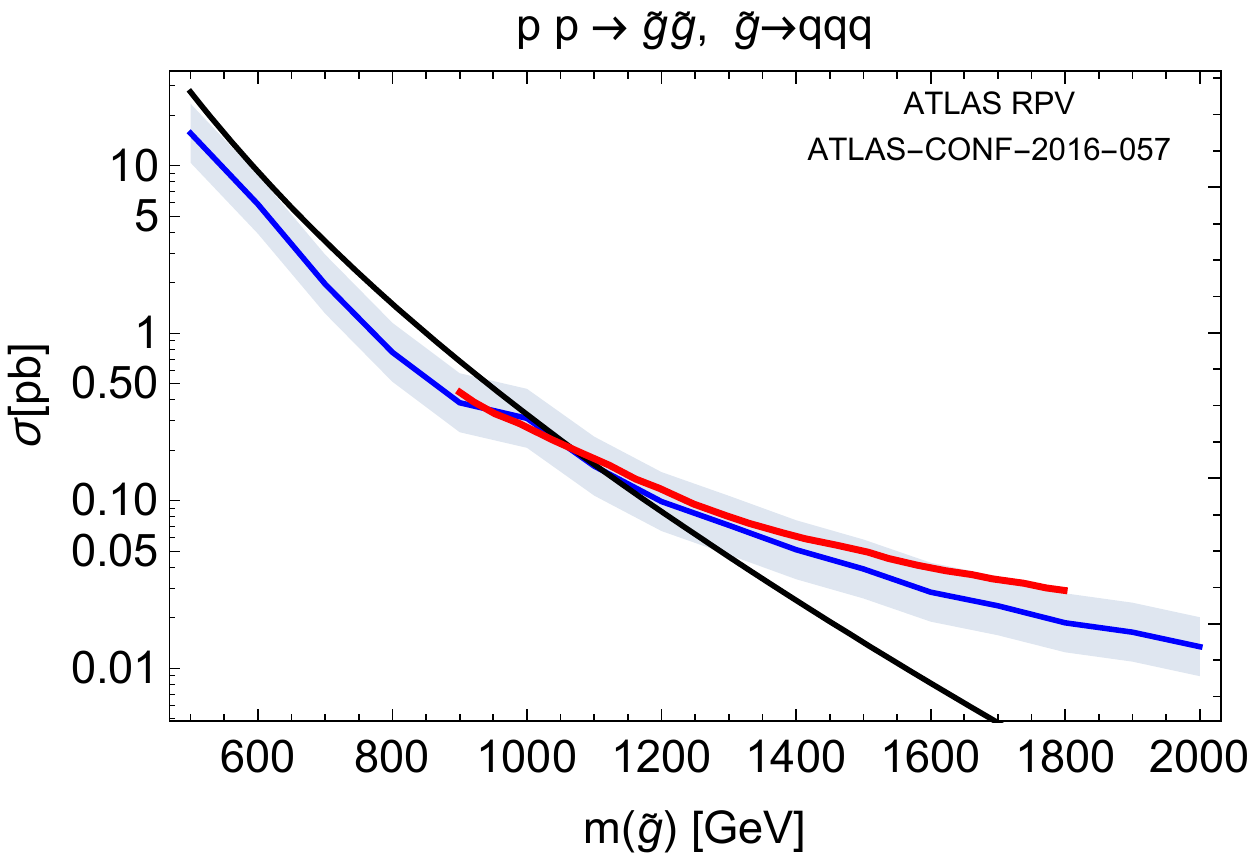}

\caption{Limits on supersymmetric simplified models from our recasted search of \cite{ATLAS:2016nij} (blue lines) with 50\% error on the number of events in the signal regions (blue shaded regions), compared to the experimental results (red lines). On the right-hand plot, the expect gluino pair production cross section is shown in black. \label{fig:ATLAS-CONF-2016-057}}
\end{figure}

We validate our recasting of the search by generating events using the supersymmetric simplified model considered by \cite{ATLAS:2016nij}: $(i)$ gluino pair production decaying to all quark pairs and a neutralino $\tilde{g} \to q\bar{q}\tilde{\chi}_1^0$ followed by neutralino decay via RPV operators into three quarks $\tilde{\chi}_1^0 \to qqq$ (with equal branching ratios to all available flavor combinations), and $(ii)$  gluino pair production decaying directly to three quarks $\tilde{g} \to qqq$.  The published limits are shown in Fig.~\ref{fig:ATLAS-CONF-2016-057} with the results of our recasted search on simulated data, along with a 50\% ``recasting uncertainty'' on the number of events in each signal region.

\subsection{ATLAS 2--6 Jets Plus MET}

The ATLAS note CONF-2016-078 \cite{ATLAS:2016kts} is a search for gluinos and squarks decaying to jets and missing energy, requiring between two and six jets, significant missing energy, and vetoing on leptons. Two search strategies are employed in \cite{ATLAS:2016kts}: one using a $m_{\rm eff}$ variable to separate signal and background, and a second using RJR variables \cite{Jackson:2016mfb}. For simplicity, we use the former search signal regions, which sets bounds as competitive as the latter. 

\begin{figure}[ht]
\includegraphics[width=0.5\columnwidth]{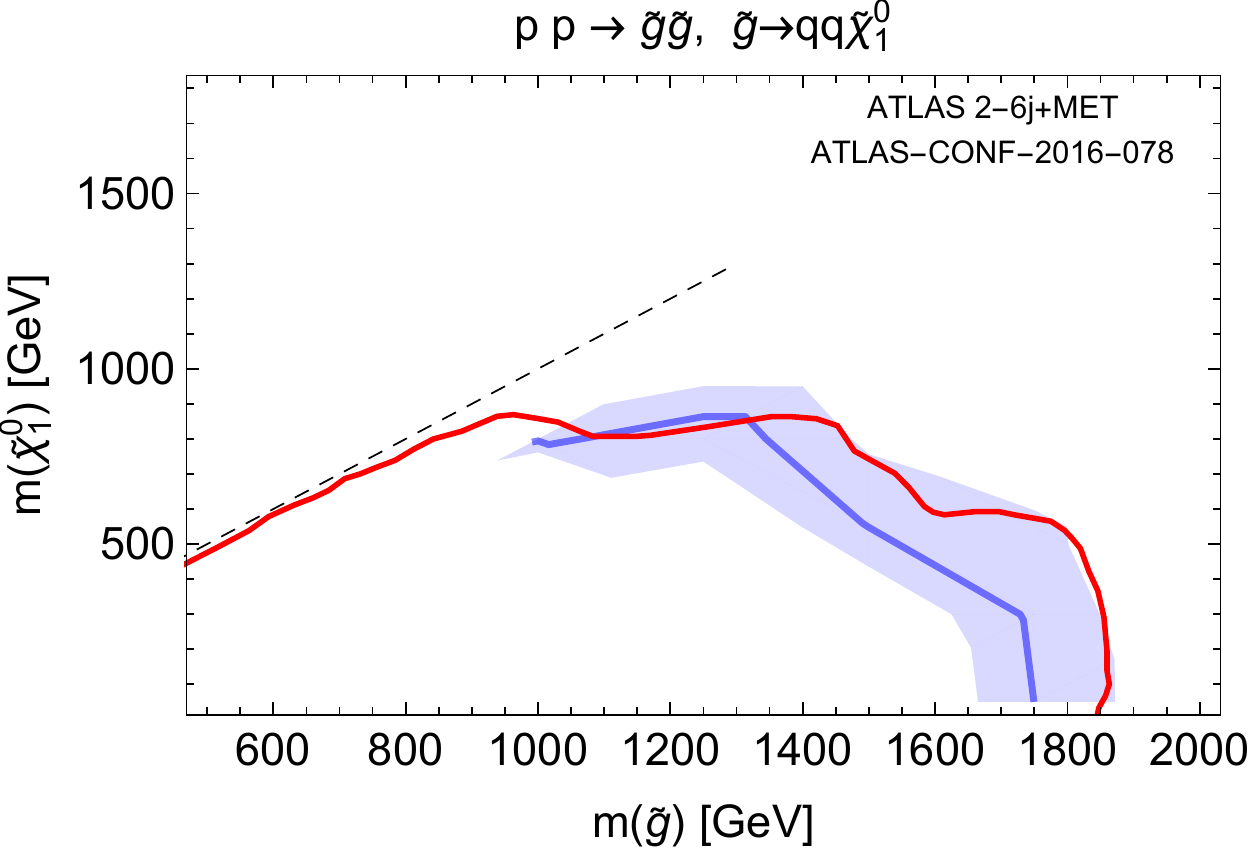}\includegraphics[width=0.5\columnwidth]{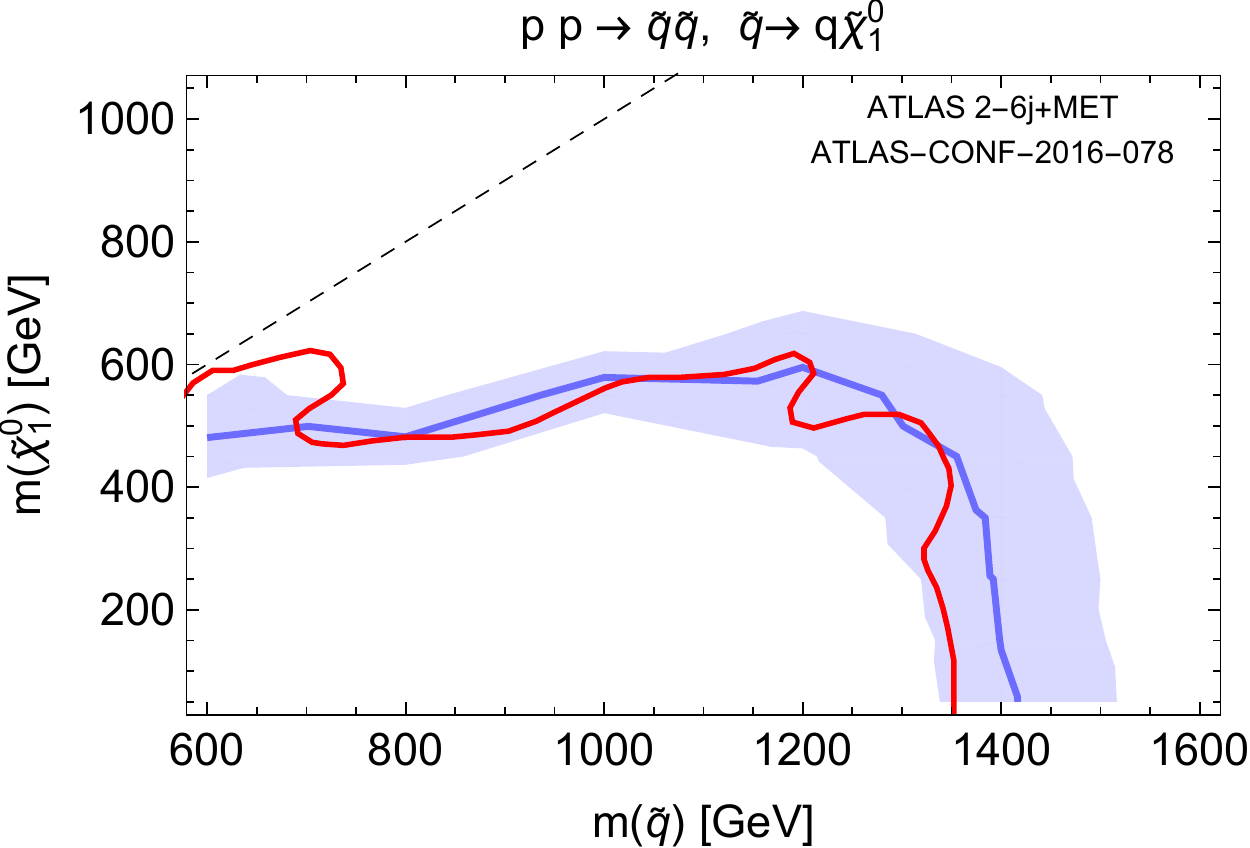}
\includegraphics[width=0.5\columnwidth]{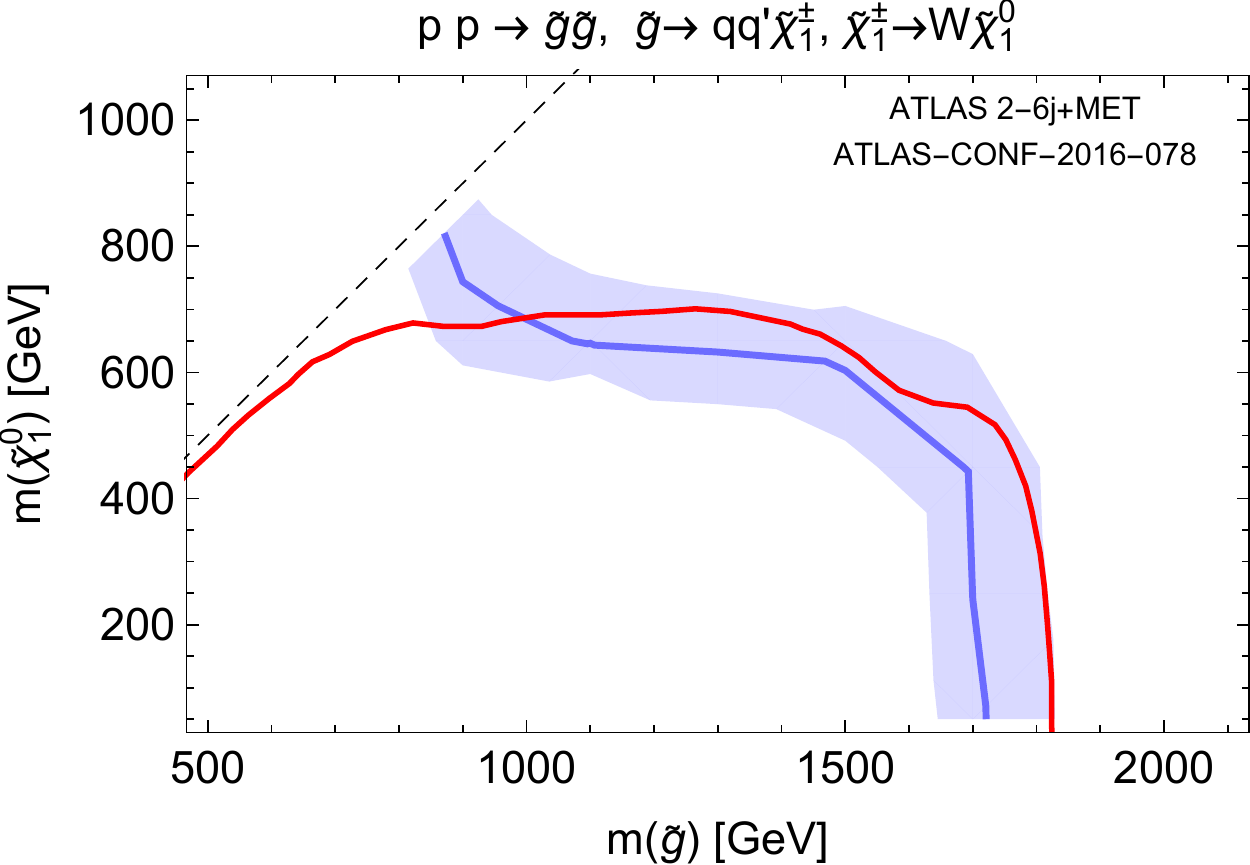}

\caption{Limits on supersymmetric simplified models from our recasted search of \cite{ATLAS:2016kts} (blue lines) with 50\% error on the number of events in the signal regions (blue shaded regions), compared to the experimental results (red lines). \label{fig:ATLAS-CONF-2016-078}}
\end{figure}

We validate our recasting of the search by generating events using the supersymmetric simplified models considered by \cite{ATLAS:2016kts}: $(i)$ gluino pair production decaying to light-flavor quarks and a neutralino $\tilde{g} \to q\bar{q}\tilde{\chi}_1^0$, $(ii)$  light-flavor squark pair production decaying to quarks and a  neutralino $\tilde{q} \to q\tilde{\chi}_1^0$, $(iii)$ and gluino pair production decaying to light-flavor quarks and a chargino, which itself decays to a neutralino and a $W$ boson: $\tilde{g} \to q\bar{q}'\tilde{\chi}_1^\pm,~\tilde{\chi}^\pm_1 \to W^\pm \tilde{\chi}_0^1$. In the latter case, the chargino mass is set to be the average of the gluino and neutralino mass. The published limits are shown in Fig.~\ref{fig:ATLAS-CONF-2016-078} with the results of our recasted search on simulated data, along with a 50\% ``recasting uncertainty'' on the number of events in each signal region.

\subsection{CMS Multi-Jet + MET}

There are three searches from the CMS Collaboration which search for gluinos and squarks decaying to jets and missing energy which were presented at ICHEP in 2016 \cite{CMS:2016mwj,CMS:2016alpha,CMS:2016xva}. We chose to work with \cite{CMS:2016mwj} (CMS-SUS-16-014), which has equivalent reach as the other two searches. This search requires at least three jets, no leptons, and significant missing transverse momentum. The full analysis uses 160 signal regions, separated by minimum jet, $b$-tagged jet, $H_T$, and $H_T^{\rm miss}$. However, this large number of signal regions can be simplified to twelve aggregated regions. In each region, we calculate the maximum number of signal events which can be accommodated at 95\% CL from the published background expectation and observation (Appendix A.5 of \cite{CMS:2016mwj}) using the CL$_{\rm s}$ method. 

\begin{figure}[ht]
\includegraphics[width=0.5\columnwidth]{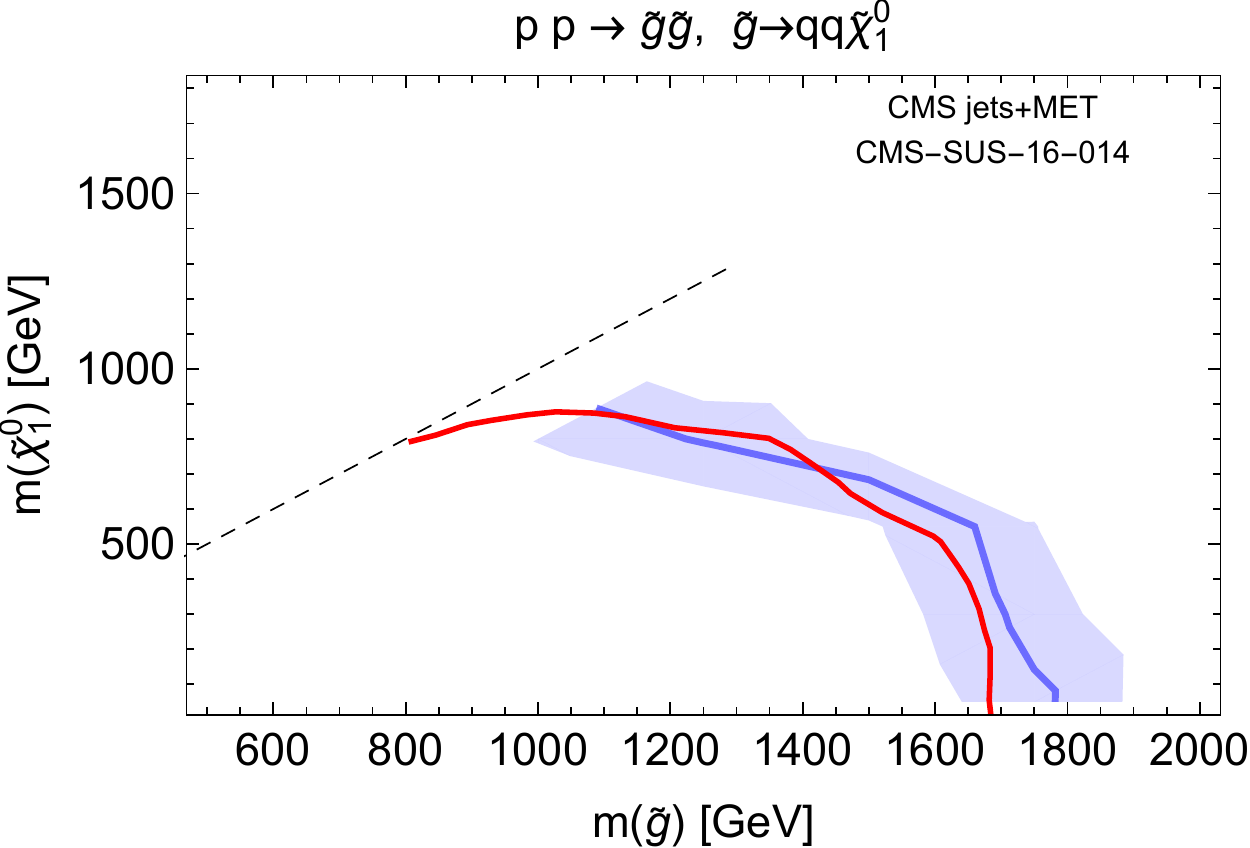}\includegraphics[width=0.5\columnwidth]{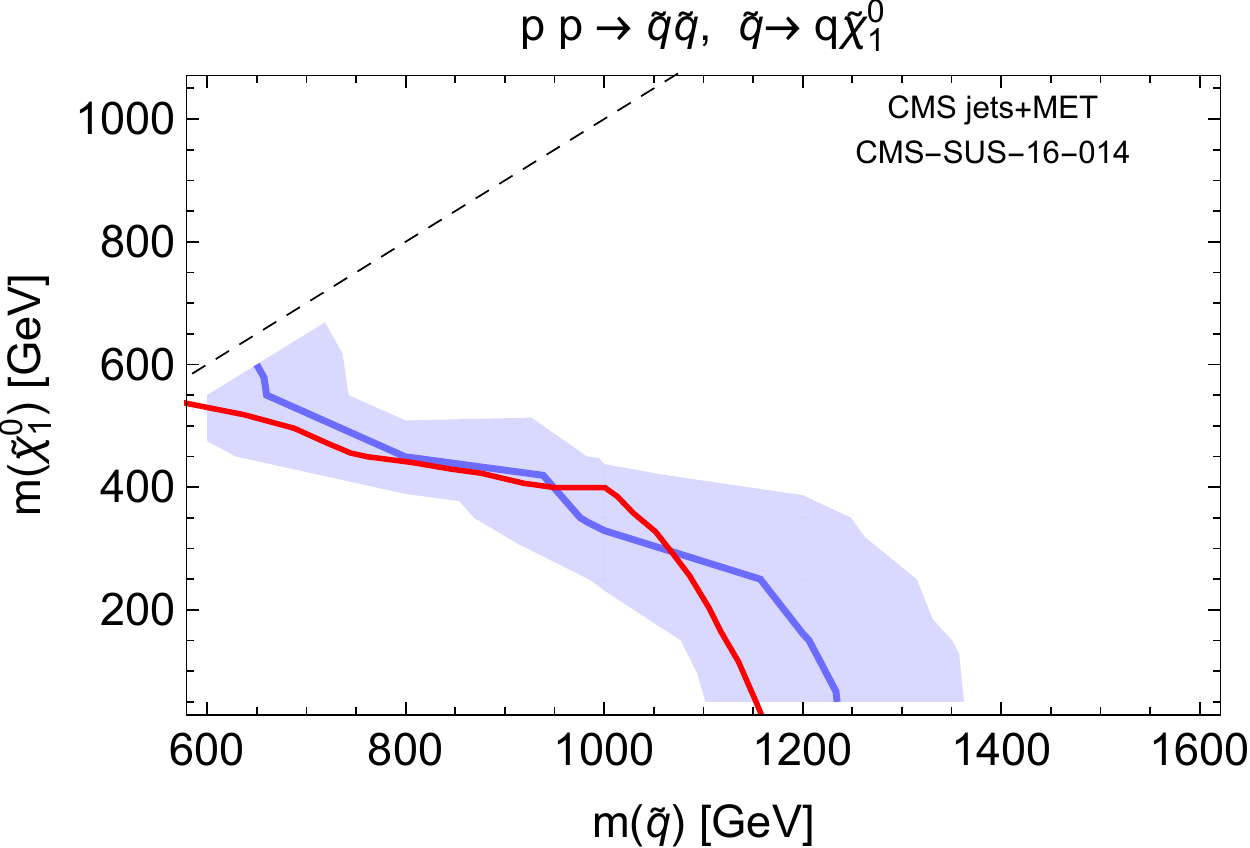}
\includegraphics[width=0.5\columnwidth]{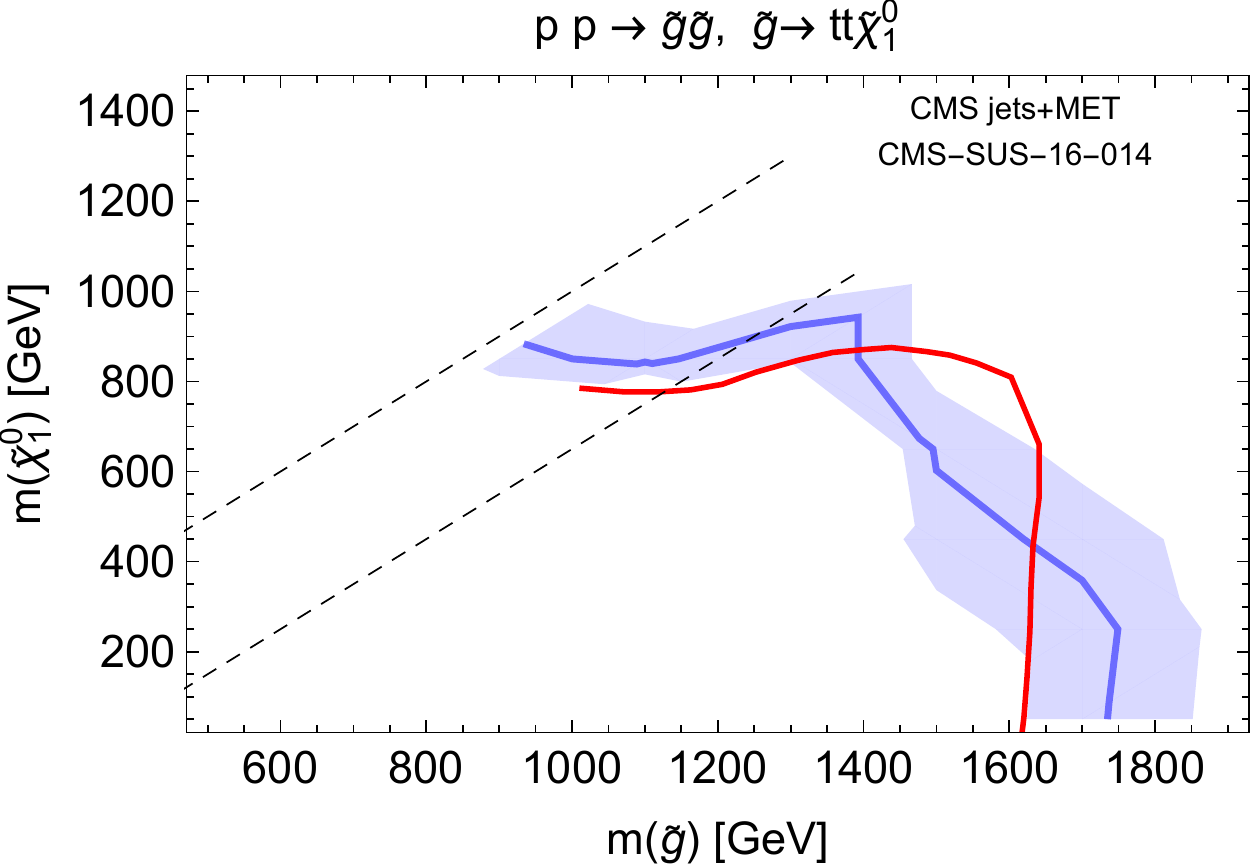}\includegraphics[width=0.5\columnwidth]{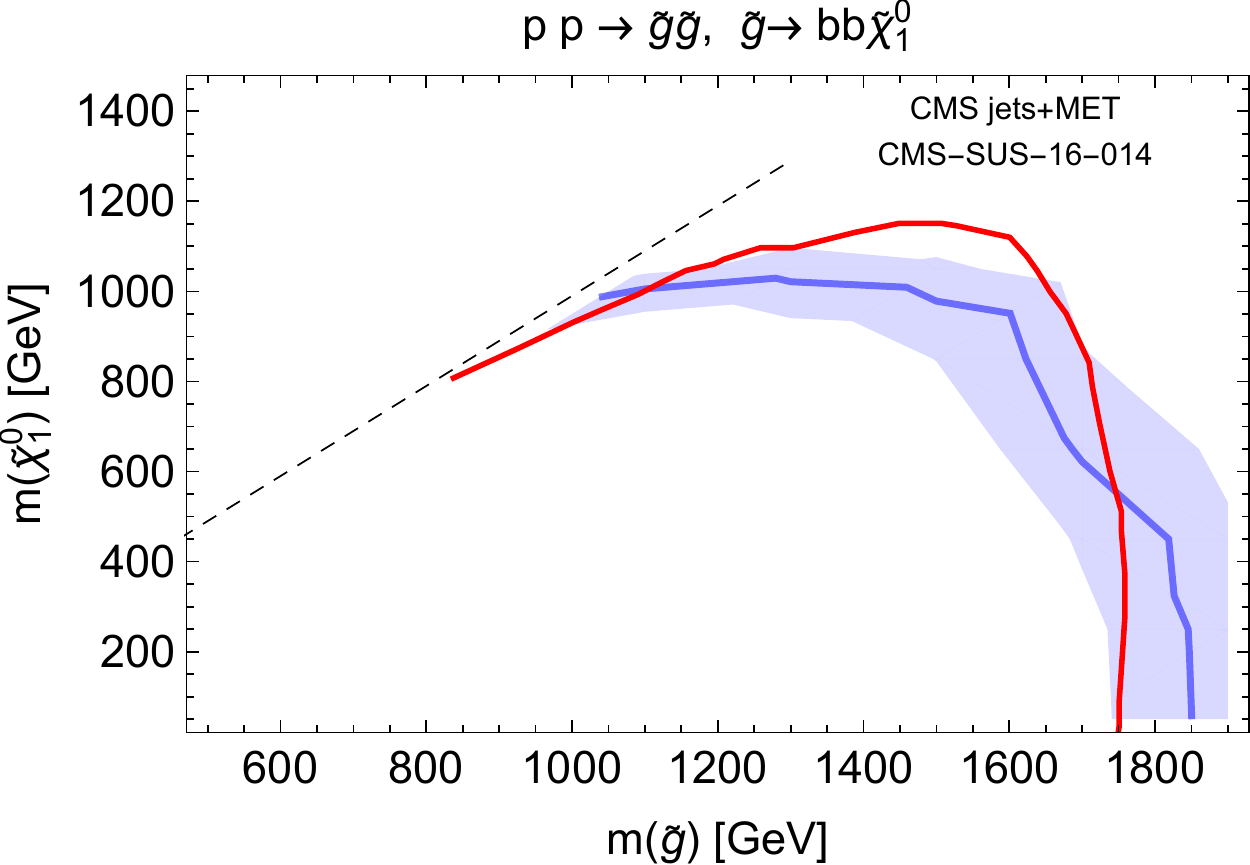}

\caption{Limits on supersymmetric simplified models from our recasted search of \cite{CMS:2016mwj} (blue lines) with 50\% error on the number of events in the signal regions (blue shaded regions), compared to the experimental results (red lines). \label{fig:CMS-SUS-16-014}}
\end{figure}

We validate our recasting of the search by generating events using the supersymmetric simplified models considered by \cite{CMS:2016mwj}: $(i)$ gluino pair production decaying to light-flavor quarks and a neutralino $\tilde{g} \to q\bar{q}\tilde{\chi}_1^0$, $(ii)$ light-flavor squark pair production decaying to quarks and a  neutralino $\tilde{q} \to q\tilde{\chi}_1^0$, $(iii)$ gluino pair production decaying to top pairs and a neutralino $\tilde{g} \to t\bar{t}\tilde{\chi}_1^0$, and $(iv)$  gluino pair production decaying to bottom pairs and a neutralino $\tilde{g} \to b\bar{b}\tilde{\chi}_1^0$. The published limits are shown in Fig.~\ref{fig:CMS-SUS-16-014} with the results of our recasted search on simulated data, along with a 50\% ``recasting uncertainty'' on the number of events in each signal region.

\subsection{ATLAS 8--10 Jets Plus MET}

The ATLAS note CONF-2016-095 \cite{ATLAS:2016kbv} (which is the direct update of the 7--10 jet search in \cite{Aad:2016jxj} with an increased luminosity of 18.2 fb$^{-1}$) is a search for gluinos decaying to jets and missing energy, requiring between eight and ten jets, some missing energy, and vetoing on leptons. ``Fat'' jets are used to discriminate over the background, in addition to the ratio $E_T^{\rm miss}/\sqrt{H_T}$.  A number of signal regions are defined, with varying jet multiplicity and requirements on the sum of the fat jet masses. These large-$R$ jets are found by reclustering the small-$R$ jets with the anti-$k_t$ algorithm and a radius $R=1.0$ in \textsc{Delphes}. Then, the sum of the masses of the reclustered jets is used to define the signal regions.

\begin{figure}[ht]
\includegraphics[width=0.48\columnwidth]{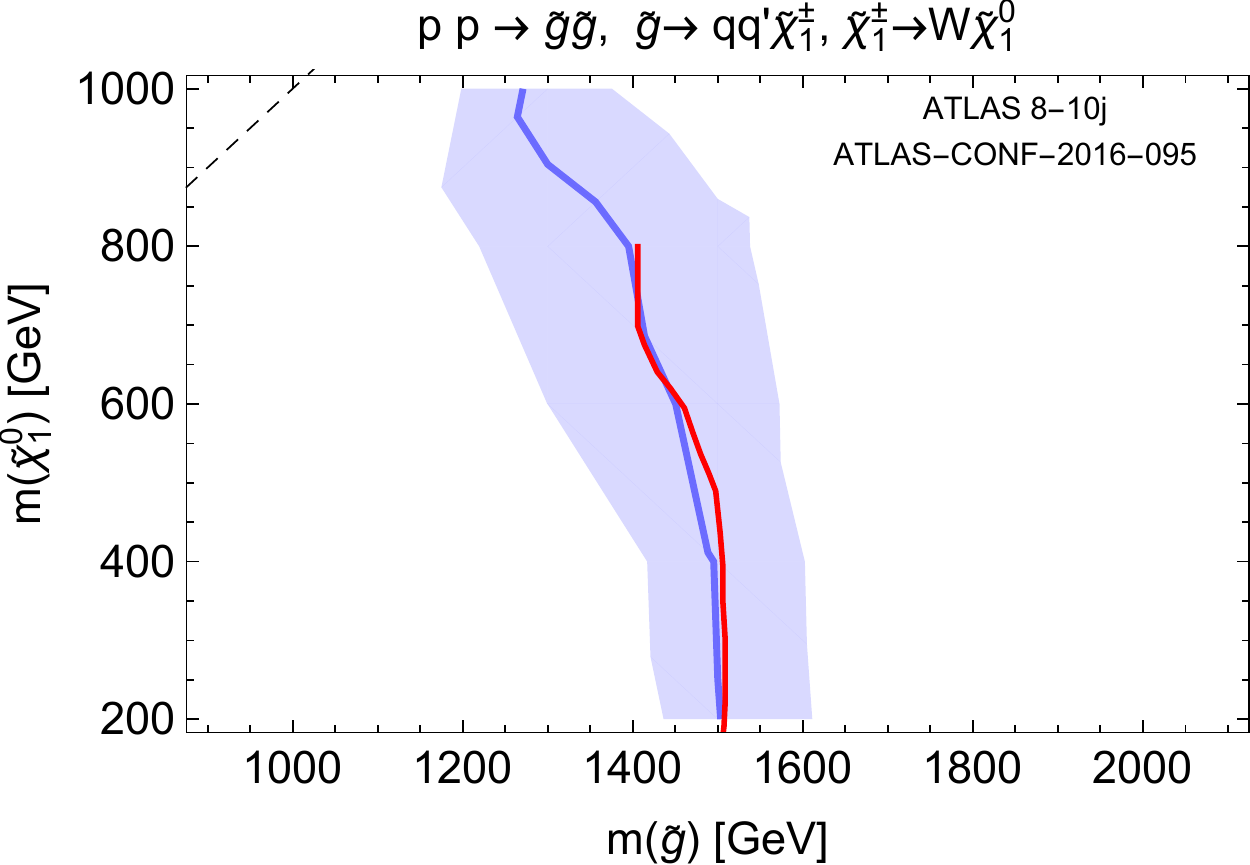}
\includegraphics[width=0.48\columnwidth]{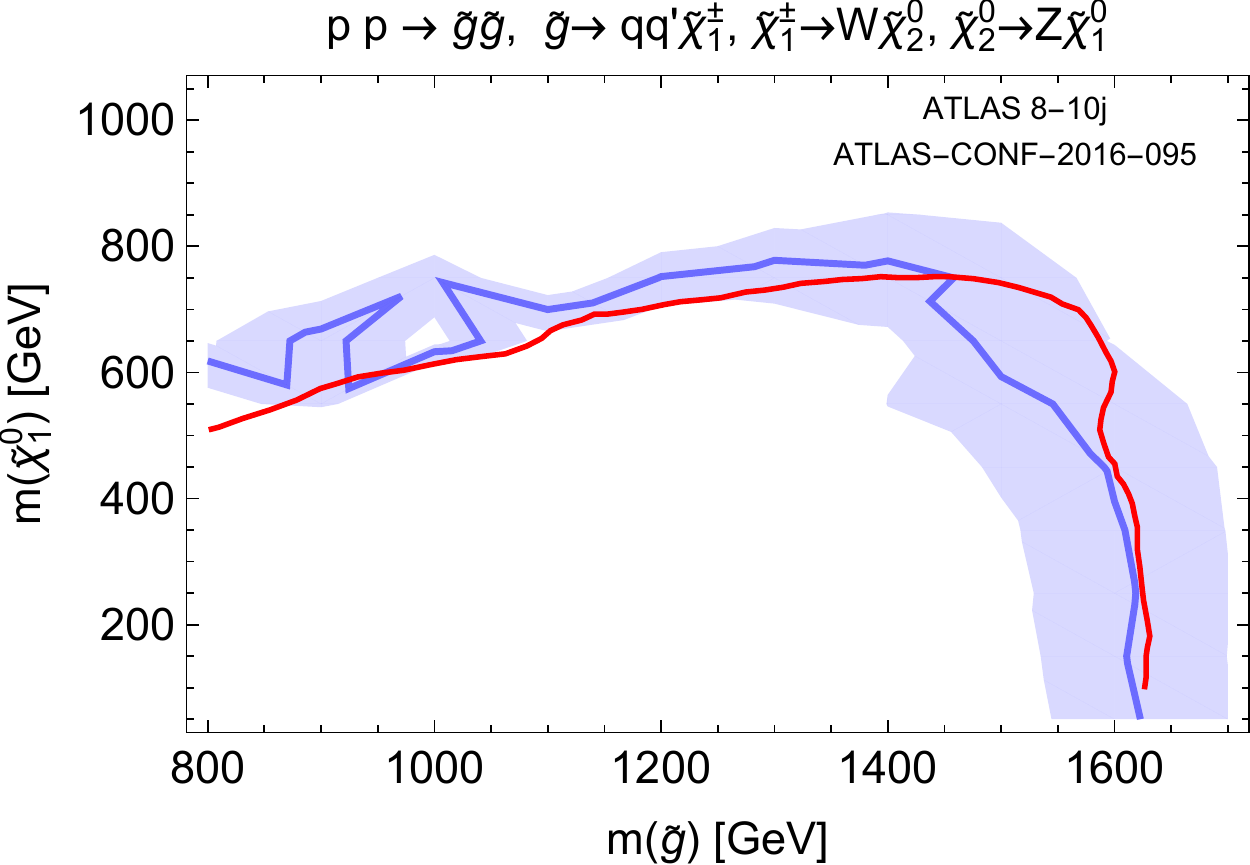}

\caption{Limits on supersymmetric simplified models from our recasted search of \cite{Aad:2016jxj} (blue line) with 50\% error on the number of events in the signal regions (blue shaded region), compared to the experimental results (red line). \label{fig:ATLAS-8-10}}
\end{figure}

We validate our recasting of the search by generating events using the supersymmetric simplified models considered by \cite{ATLAS:2016kbv}, 
$(i)$ gluino pair production decaying to light-flavor quarks and a chargino, which itself decays to a neutralino and a $W$ boson, $\tilde{g} \to q\bar{q}'\tilde{\chi}_1^\pm,~\tilde{\chi}^\pm_1 \to W^\pm \tilde{\chi}_0^1$, with the chargino mass set to be the average of the gluino and neutralino mass, and $(ii)$ gluino pair production decaying to light-flavor quarks and a chargino, which then  decays to a $W$ boson and a neutralino $\tilde \chi^0_2$, followed by the neutralino decay to a $Z$ boson and the lightest neutralino  $\tilde \chi^0_1$, $\tilde{g} \to q\bar{q}'\tilde{\chi}_1^\pm,~\tilde{\chi}^\pm_1 \to W^\pm \tilde{\chi}_0^2, \tilde{\chi}_0^2\to Z \tilde{\chi}_0^1$. 
The published limits are shown in Fig.~\ref{fig:ATLAS-8-10} with the results of our recasted search on simulated data, along with a 50\% ``recasting uncertainty'' on the number of events in each signal region.

\subsection{ATLAS Lepton Plus Many Jets}
The ATLAS note CONF-2016-094 \cite{ATLAS:2016mnt} is a search for gluinos decaying to top-rich final states and little missing transverse momentum, requiring one lepton and multiple jets in the final state. A number of signal regions are defined, separated by number of jets and $b$-tagged jets.

\begin{figure}[ht]
\includegraphics[width=0.48\columnwidth]{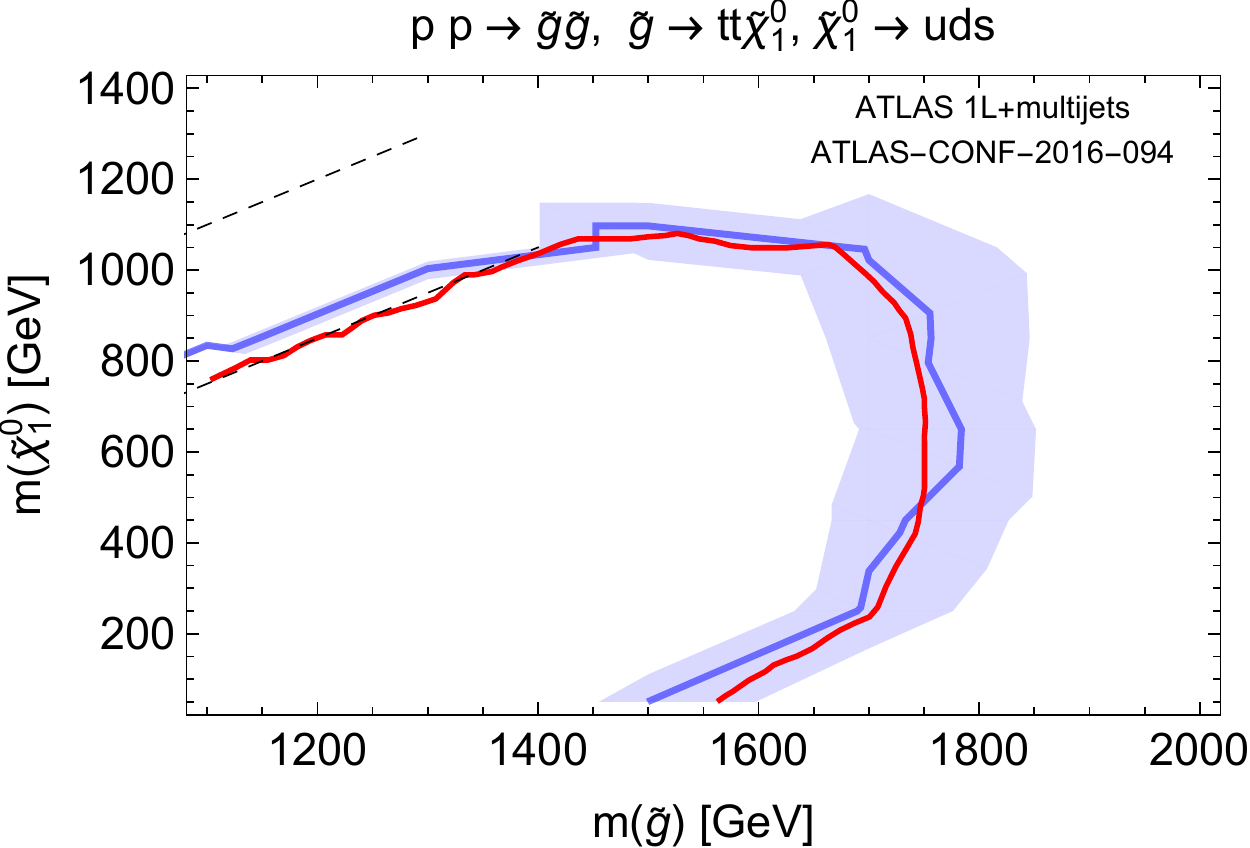}

\caption{Limits on supersymmetric simplified models from our recasted search of \cite{ATLAS:2016mnt} (blue line) with 50\% error on the number of events in the signal regions (blue shaded region), compared to the experimental results (red line). \label{fig:ATLAS-CONF-2016-094}}
\end{figure}

We validate our recasting of the search by generating events using the supersymmetric simplified model considered by \cite{ATLAS:2016mnt}: gluino pair production decaying to top quarks and a neutralino, which itself decays via RPV to three light quarks: $\tilde{g} \to t\bar{t}'\tilde{\chi}_1^0,~\tilde{\chi}^0_1 \to uds$. The published limits are shown in Fig.~\ref{fig:ATLAS-CONF-2016-094} with the results of our recasted search on simulated data, along with a 50\% ``recasting uncertainty'' on the number of events in each signal region.

\end{document}